\documentclass[graybox]{svmult}

\usepackage{type1cm}        
\usepackage{makeidx}         
\usepackage{graphicx}        
\usepackage{multicol}        
\usepackage[bottom]{footmisc}
\usepackage{amssymb}
\usepackage{graphicx,setspace}
\usepackage{amsmath}
\usepackage{epsfig}
\usepackage{amssymb}

\makeindex             

\begin{document}

\title{Understanding Hawking radiation from simple models of atomic Bose-Einstein condensates}
\titlerunning{Simple models of Hawking radiation in atomic BECs}
\author{R. Balbinot, I. Carusotto, A. Fabbri, C. Mayoral, and A. Recati}
\institute{Roberto Balbinot \at Dipartimento di Fisica dell'Universit\`a di Bologna and INFN sezione di Bologna, Via Irnerio 46, 40126 Bologna, Italy \\ \email{balbinot@bo.infn.it}
\and Iacopo Carusotto and Alessio Recati \at INO-CNR BEC Center and Dipartimento di Fisica, Universit\`a di Trento, via Sommarive 14, 38123 Povo, Trento, Italy\\ \email{carusott@science.unitn.it, recati@science.unitn.it}
\and Alessandro Fabbri and Carlos Mayoral \at Departamento de F\'isica Te\'orica and IFIC, Universidad de Valencia-CSIC, C. Dr. Moliner 50, 46100 Burjassot, Spain \\ \email{afabbri@ific.uv.es, carlosmsaenz@gmail.com}
}
%
%
\maketitle


\abstract{This chapter is an introduction to the Bogoliubov theory  of dilute Bose condensates as applied to the study of the spontaneous emission of phonons in a stationary condensate flowing at supersonic speeds. This emission process is a condensed-matter analog of Hawking radiation from astrophysical black holes but is derived here from a microscopic quantum theory of the condensate without any use of the analogy with gravitational systems. To facilitate physical understanding of the basic concepts, a simple one-dimensional geometry with a stepwise homogenous flow is considered which allows for a fully analytical treatment.}

\section{Introduction}
\label{sec:1}


One of the most spectacular predictions of Einstein's General Relativity is the existence of Black Holes (BHs), mysterious objects whose gravitational field is so strong that not even light can escape from them but remains trapped inside a horizon. According to the standard view, BHs are formed by the collapse of massive stars ($M>3M_{Sun}$) at the end of their thermonuclear evolution when the internal pressure is no longer able to balance the gravitational self attraction of the star. Furthermore supermassive BHs ($M>10M_{Sun}$) are supposed to constitute the inner core of active galaxies.

As no light can escape from them, BHs are expected to be really ``black" objects. In particular, their observational evidence can only be indirect: typically, the presence of a black hole is deduced by observing the behavior of matter (typically hot gas) orbiting outside the horizon. A hypothetical isolated BH (i.e. a BH immersed into vacuum) would not manifest its presence except for his gravitational field, which after a short transient time becomes stationary (even static if there is no angular momentum).

In 1974 Hawking showed \cite{hawking} that this common belief is incorrect. If one takes into account Quantum Mechanics, static and stationary BHs are no longer ``black", but rather emit a steady radiation flux with a thermal spectrum at a temperature given, simply speaking, by the gradient of the gravitational potential at the horizon. This intrinsically quantum mechanical process is triggered by the formation of the horizon and proceeds via the conversion of vacuum fluctuations \index{vacuum fluctuations} into on shell particles. This effect is a universal feature of BHs, completely independent of the details of the BH formation.

In spite of the interest that this fascinating effect has raised in a wide audience, no experimental evidence is yet available in support of this amazing theoretical prediction. Since the emission temperature scales as the inverse of the BH mass ($T\sim 10^{-7}\,\textrm{K}$ for a solar mass BH), the expected Hawking signal is in fact many order of magnitudes below the $2.7\,\textrm{K}$ of the cosmic microwave background.
As a result, the Hawking radiation \index{Hawking radiation} by BHs appears to be a completely irrelevant process in any realistic astrophysical situation, with no hope to be detected in the sky. This situation is rather frustrating, since the conceptual relevance of Hawking discovery is extremely profound: the existence of Hawking radiation Hawking radiation allows such a beautiful synthesis between gravity and thermodynamics that it cannot be just an accident; many people indeed regard Hawking result as a milestone of the still to be discovered quantum theory of gravity.

After almost 40 years of research on BHs, the attitude nowadays appears a bit different and more promising on the experimental side.
In particular, it was realized that the Hawking emission process is not at all peculiar to gravitational physics: its ``kinematical" rather than ``dynamical" nature makes it manifest itself in different physical contexts. This way of looking at the Hawking effect has its origin in a paper by Unruh in 1981 \cite{unruh} where a steady emission of thermal phonons was predicted to appear in any fluid in stationary flow turning supersonic: the basic process underlying this phonon emission is completely identical to the one discussed by Hawking for the gravitational BH, in the sense that the mathematical equations describing it are exactly the same as the ones describing Hawking radiation  from gravitational BHs. The reason for this amazing and unexpected ``analogy" is that the equation describing the propagation of long wavelength sound waves in a moving fluid can be recast in terms of a massless scalar field propagating in a curved spacetime with a suitably chosen ``acoustic metric''. In particular, the point where a sub-sonic flow turns supersonic plays the role of an ``acoustic horizon" \index{acoustic horizon}, since sound waves in the supersonic region are no longer able to propagate upstream. As it happens to light inside a BH, sound waves are trapped inside the sonic horizon of the \index{acoustic black hole} ``acoustic black hole": upon quantization, it is then straightforward to expect the emission of Hawking radiation  by the horizon. Nowadays, we know that this analogy with gravitational systems is not limited to fluids but can be developed for many other condensed matter and optical systems~\cite{livrel}. Unlike gravitational BHs, these condensed matter analog models often possess a well understood quantum description at the microscopic level, which allows for a complete control of their physics. This is the case, in particular, of atomic Bose-Einstein condensates \index{Bose-Einstein condensates} which are the subject of the present chapter.

The relevance of the analogy is therefore twofold. At one hand, one can concretely consider investigating the actual existence of Hawking radiation  using table top experiments with a complete control of the physical system.
On the other hand, the detailed knowledge of the underlying microscopic quantum theory of these systems allows us to address a very delicate point in the theory of Hawking radiation  and possibly to eliminate some intrinsic inconsistencies of its standard derivation.

In the absence of a complete and self consistent quantum theory of gravity, one typically adopts a semiclassical framework where gravity is treated classically according to General Relativity, whereas light and matter fields propagating on top of the curved space time are quantized. This is the so called Quantum Field Theory in Curved Space \cite{birdav}. One expects this scheme to provide a sufficiently accurate description of the gravity-matter systems for scales sufficiently large as compared to the fundamental quantum scale for gravity, the so-called Planck scale equal to $10^{-33}\,$cm or $10^{19}\,$GeV. Approaching this Planck scale, one can reasonably expect that this semiclassical description becomes inaccurate and has to be replaced by a (yet to be discovered) complete theory of quantum gravity.

Now because of the infinite (exponential) redshift suffered by the Hawking phonons in their journey from the near horizon region to infinity, a given mode of Hawking radiation  measured at time $t$ with frequency $\nu$ far from the BH appears to have had a frequency $\nu'=\nu\,e^{{ct}/{2R}}$ near the BH horizon ($R$ is the radius), which rapidly overhangs Planck energy: this feature makes the derivation clearly inconsistent and casts serious doubts on the very existence of Hawking BH radiation. This is the so called transplanckian problem \cite{transpl}.

The same kind of argument can be repeated also for Hawking like radiation in condensed matter systems: because of the infinite Doppler shift at the sonic horizon, the modes responsible for Hawking like radiation oscillate near the horizon at a wavelength much smaller than the intermolecular or interatomic spacing, which makes the hydrodynamical long wavelength approximation inconsistent. On this basis, it would therefore be difficult to rule out the possibility that Hawking radiation be an artifact due to an illegitimate extrapolation of the long wavelength approximation, that is a spurious outcome without any physical reality.

In this perspective, analogue condensed matter systems provide a new angle from which the transplanckian problem of Hawking radiation  can be attacked: as they possess a detailed and well understood microscopic quantum description, the question of the existence of Hawking radiation  can be investigated from first principles, without any use of the hydrodynamical approximation and hence of concepts borrowed from the gravitational analogy \index{gravitational analogy} like effective metric, horizon, etc. So far, most of the work in this direction has been performed using atomic BECs, but the fully positive answer coming from these studies appears to hold under very general assumptions: Hawking radiation is indeed a real physical phenomenon!

A closer look at the spectral and coherence properties of the predicted Hawking radiation match the original expectation that, if the transition is sufficiently smooth with respect to the microscopic scales of the fluid, the Hawking emission of Bogoliubov phonons is thermal at a temperature proportional to the gradient of the flow potential at the sonic point~\cite{machpare}. In addition, several novel interesting features have pointed out in regimes beyond the hydrodynamical approximation as well as in different configurations, e.g. white holes (the time-reversed of a black hole)~\cite{undu} and the so-called black-hole lasers (a pair of adjacent black and white hole horizons)~\cite{finazzi_laser}.

Simultaneously to these theoretical and conceptual advances, a great effort has been devoted in the last years to the identification of the most promising physical systems where to experimentally investigate the analogue Hawking radiation.

Having established that the Hawking effect exists, one can start to think at the best experimental setting to reveal it. There are many systems proposed at this end, like ultracold atoms, optical systems, water tank experiments and others.
At the moment, experiments with water tanks \cite{unwein} have detected the classical counterpart of Hawking emission in flows showing white hole horizons: stimulated emission by the Hawking mechanism is probed by sending a classical incident wavepacket of surface waves against the horizon. Unfortunately, these experiments at room temperature do not appear suitable to investigate the very quantum phenomenon of Hawking radiation, that is the conversion of zero-point fluctuations into observable quanta by the horizon.
An observation of Hawking radiation  from laser pulses propagating in nonlinear optical media has been recently reported~\cite{faccio}, but this result is still object of intense discussion in the community \cite{scun}.

The main experimental difficulty in the quest for analog Hawking radiation in condensed matter systems is the extremely weak intensity of the signal in realistic systems, which makes it to be easily covered by competing effects like the thermal emission due to the non zero temperature of the systems as well as quantum noises. In this respect, atomic gases appear as most promising systems~\cite{garay}, as they combine a variety of tools for the manipulation of the diagnostic of the system down to microscopic level, to the possibility of cooling the system to very low temperatures where the zero point quantum fluctuations start playing an important role. Still, even in these systems temperatures lower than the expected Hawking temperature of the order of $10\ nK$ are hardly reached, and further difficulty comes from the detection of the Hawking phonons emitted from the horizon.

A major breakthrough that appears to bypass both these problems was proposed by us in 2008 \cite{lettera} and is based on the use of density correlations \index{density correlations}, a modern powerful tool to investigate microscopic properties of strongly correlated atomic gases and in particular of their elementary excitations. Taking advantage of the fact that the Hawking radiation  consists of correlated pairs of quanta emitted in opposite directions from the horizon, a characteristic signal will appear in the density-density correlation function for points situated on opposite sides with respect to the horizon. This unique signature was made quantitative using gravitational physics methods and then numerically confirmed by {\em ab initio} simulations of the condensate dynamics based on a microscopic description of their collective properties~\cite{numerics}. As a result, it appears to be an ideal tool to isolate the Hawking radiation  signal from the background of competing processes and of experimental noise even at non-zero temperatures. Of course, a similar strategy would be clearly impossible in astrophysical black holes, as no access is possible to the region beyond the horizon.

In this paper we shall use standard tools of the theory of dilute Bose gas to show in a rather pedagogical way how Hawking radiation  emerges in an atomic BEC and to explain its features on a simple and analytically tractable toy model. Our treatment, as we shall see, resembles very much learning elementary Quantum Mechanics using one dimensional Schr\"odinger equation with square potentials. Most of the material presented here was originally published in Refs.\cite{rpc} and \cite{cfr}.

\section{The theory of dilute Bose-Einstein condensates \index{Bose-Einstein condensates} in a nutshell
}
\label{sec:2}



%
%

In this section we give a brief and rapid introduction to the theory of BECs. In particular, we shall review the Gross-Pitaevskii equation \index{Gross-Pitaevskii equation} describing the dynamics of the condensate at mean field level and the Bogoliubov \index{Bogoliubov theory} description of quantum fluctuations of top of it. More details can be found in textbooks~\cite{SSLP} and in dedicated reviews~\cite{castin}.

Bose-Einstein condensation is characterized by the accumulation of a macroscopic fraction of the particles into a single quantum state. To achieve such a quantum degeneracy very low temperatures are required (on the order of $T=100\ nK$ for the typical densities of ultracold atomic gases in magnetic or optical traps), where particles are no longer distinguishable and their Bose statistics start being relevant.

\subsection{The Gross-Pitaevskii equation  and the Bogoliubov theory}

The model Hamiltonian describing a many-body system composed of $N$ interacting bosons confined in an external potential $V_{ext}(\vec{x})$ can be written in a second quantized formalism as:
\begin{equation}
\hat H=\int d^3x\left[ \hat\Psi^{\dagger}\left( -\frac{\hbar^2}{2m}\nabla^2 + V_{ext}\right)\hat\Psi + \frac{g}{2}\hat\Psi^{\dagger}\hat\Psi^{\dagger}\hat\Psi\hat\Psi \right]
\label{eq:H}
\end{equation}
where $\hat\Psi(t,\vec x)$ is the field operator which annihilates an atom at position $\vec x$ and obeys standard bosonic equal time commutation rules
\begin{equation}\label{etc}
[\hat \Psi (\vec x), \hat \Psi^{\dagger}(\vec x')]=\delta^3(\vec x- \vec x').
\end{equation}
The model Hamiltonian (\ref{eq:H}) is generally used within the dilute gas approximation where the two body interatomic potential can be approximated by a local term $V(x-x')=g\delta^3(\vec x- \vec x')$ with an effective coupling constant $g$ related to the atom-atom scattering length $a$ by $g=4\pi\hbar^2a/m$.

At sufficiently low temperatures well below the Bose-Einstein condensation temperature, a macroscopic fraction of atoms is accumulated into the single one-particle state of lowest energy, described by the macroscopic wavefunction $\Psi_0(\vec x)$.
The time evolution of the macroscopic wavefunction in response of some excitation (e.g. a temporal variation of the confining potential $V_{ext}$) is described by the {\em Gross-Pitaevski equation \index{Gross-Pitaevski equation}}
\begin{equation}\label{gp}
i\hbar\frac{\partial \Psi }{\partial t} = \left(-\frac{\hbar^2}{2m}\vec \nabla^2 + V_{ext} + g |\Psi|^2 \right)\,\Psi:
\end{equation}
whose form can be heuristically derived by performing a mean-field approximation $\hat\Psi\rightarrow \Psi_0$ in the Heisenberg equation
\begin{equation}
i\hbar \frac{\partial \hat\Psi(t,\vec x)}{\partial t}=\left[\hat \Psi(t,\vec x), \hat H\right]
\end{equation}
 for the time-evolution of the atomic quantum field operator $\hat\Psi$. The ground state wavefunction naturally emerges as the lowest-energy steady-state $\Psi_0(\vec x)$ of the Gross-Pitaevskii equation  and oscillates at a frequency $\mu/\hbar$.

Small fluctuations around the mean-field can be studied within the so-called Bogoliubov approximation, where the bosonic field operator $\hat \Psi$ is written as the sum of a mean-field classical field plus quantum fluctuations. In its usual formulation to describe weakly excited condensates, one takes a steady state $\Psi_0$ as the mean-field,
\begin{equation}
\hat\Psi(t,\vec x)=\Psi_0(\vec x)\,[1+\hat \phi(t,\vec x)]\,e^{-i \mu t/\hbar}.
\label{eq:Bogo_exp}
\end{equation}
The field operator $\hat \phi$ describing fluctuations then satisfies the Bogoliubov-de Gennes \index{Bogoliubov-de Gennes} (BdG) equation
\begin{equation}\label{bdg}
i\hbar \frac{ \partial \hat \phi}{dt}= - \left( \frac{\hbar^2 \vec \nabla^2}{2m} + \frac{\hbar^2}{m}\frac{\vec \nabla \Psi_0 }{\Psi_0} \vec \nabla\right)\hat\phi +ng (\hat\phi + \hat\phi^{\dagger}),
\end{equation}
where $n=|\Psi_0|^2$.
The next subsections will be devoted to a rewriting of the BdG equation in terms of a curved space-time with an effective metric determined by the spatial profiles of the local speed of sound $c=\sqrt{ng/m}$ and of the local flow velocity $\vec v_0$.

\subsection{Analogue gravity in atomic BECs}
\label{sec:3}

We stop for the moment the formal development of BEC theory and show how a different parametrization of the field operator leads to a reinterpretation of the above equations in a hydrodynamical language and then to the gravitational analogy \index{gravitational analogy} \cite{livrel}.

Using the so called density-phase representation of the condensate wavefunction $\Psi_0=\sqrt{n}\,e^{i\theta}$, the Gross-Pitaevskii equation \index{Gross-Pitaevskii equation} (\ref{gp}) can be rewritten as a pair of real equations,
\begin{eqnarray}
\label{conteq}
&&\partial_t n + \nabla (n\vec v)=0\ ,\\ \label{quasieuler}
&&\hbar \partial_t\theta=-\frac{\hbar^2}{2m}(\nabla\theta)^2-gn-V_{ext}-V_{q}:
\end{eqnarray}
the former equation Eq. (\ref{conteq}) is the continuity equation with an irrotational~\footnote{From the definition of the velocity field $\vec v_0$, it is immediate to see that the vorticity in the condensate can only appear at points where the density vanishes.}
condensate velocity $\vec v_0=\hbar\,\nabla\theta / m$. The latter is analogous to Euler equation for an irrotational inviscid fluid, with an additional ``quantum pressure" term $V_q(\vec x)$
\begin{equation}
V_q \equiv -\frac{\hbar^2}{2m}\frac{\nabla^2\sqrt{n}}{\sqrt{n}}
\end{equation}
describing a kind of stiffness of the macroscopic wavefunction.

In this density-phase representation, the Bogoliubov expression (\ref{eq:Bogo_exp}) of the field operator is rewritten as
\begin{equation}
\hat\Psi=\sqrt{n+\hat n_1}\,e^{i(\theta+\hat \theta_1)}\simeq \Psi_0\left(1+ \frac{\hat n_1}{2n}+ i\hat\theta_1 \right)
\end{equation}
and the Bogoliubov equation (\ref{bdg}) reduce to a pair of equations of motion for the fluctuations in the density $\hat n_1$ and in the phase ($\hat\theta_1$) in the form
\begin{eqnarray}\label{bodg1}
&&\hbar\partial_t \hat\theta_1=-\hbar \vec v_0\nabla \hat \theta_1 - \frac{mc^2}{n}\hat n_1
 + \frac{mc^2}{4n}\xi^2 \nabla [ n\nabla(\frac{\hat n_1}{n})]=0
 \ ,\\ \label{bodg2}
&&\partial_t\hat n_1=-\nabla (\vec v_0 \hat n_1+ \frac{\hbar n}{m}\nabla\theta_1 ).
\end{eqnarray}
Here, a fundamental length scale is set by the so-called {\em healing length} defined as $\xi\equiv \hbar/mc$ in terms of the local speed of sound $c=\sqrt{ng/m}$.

If one is probing the system on length scales much larger than $\xi$ (the so-called {\em hydrodynamic approximation} \index{hydrodynamic approximation}), the last term in eq. (\ref{bodg1}) can be neglected. As a result, the density fluctuations can be decoupled as
\begin{equation}
\hat n_1= -\frac{\hbar n}{mc^2}\left[ \vec v_0\nabla\hat\theta_1 +\partial_t\hat\theta_1\right].
\end{equation}
When this form is inserted in eq. (\ref{bodg2}), the equation of motion for the phase perturbation
\begin{equation}\label{eqfase}
-(\partial_t+\nabla\vec v_0)\frac{n}{mc^2}(\partial_t+\vec v_0\nabla)\theta_1 + \nabla\frac{n}{m}\nabla\theta_1=0
\end{equation}
can be rewritten in a matrix form
\begin{equation}
\partial_{\mu}(f^{\mu\nu}\partial_{\nu}\hat\theta_1)=0
\end{equation}
where the matrix elements $f^{\mu\nu}$ are defined as
\begin{equation}
f^{00}=-\frac{n}{c^2}, \; f^{0i}=f^{i0}=-\frac{n}{c^2}v_0^i, \; f^{ij}=\frac{n}{c^2}(c^2\delta^{ij}-v_0^iv_0^j)
\end{equation}
in terms of the condensate density $n$ and local velocity $\vec v_0$. Greek indices $\mu,\nu=0,1,2,3$ indicate 4-dimensional objects, while latin ones $i=1,2,3$ indicate the space coordinates.

Now in any Lorentzian manifold the curved space scalar d'Alembertian operator can be written as
\begin{equation}
\Box=\frac{1}{\sqrt{-g}}\partial_\mu (\sqrt{-g}g^{\mu\nu}\partial_\nu )
\end{equation}
where $g$ is the metric, $g^{\mu\nu}$ its inverse and $g=det(g_{\mu\nu})$.
Keeping this in mind, Eq. (\ref{eqfase}) for the condensate phase dynamics can be rewritten in the form of a curved space wave equation
\begin{equation}\label{kgeq}
\Box \theta_1=0\ ,
\end{equation}
provided one identifies
\begin{equation}
\sqrt{-g}\;g^{\mu\nu}\equiv f^{\mu\nu}\ ,
\end{equation}
which can be inverted leading to the effective metric
\begin{equation}\label{acm}
g_{\mu\nu}=\frac{n}{mc}\left(
  \begin{array}{cccc}
   -(c^2-\vec v_0^2) & -v_0^i \\
     -v_0^j & \delta_{ij} \\
  \end{array}\right).
\end{equation}

In summary, we have shown that under the hydrodynamical approximation, the equation of motion for the phase fluctuation in a BEC can be rewritten in terms of a  Klein-Gordon equation for a massless scalar field propagating in a fictitious space-time described by the metric $g_{\mu\nu}$  defined by Eq.(\ref{acm}). This is the core of the gravitational analogy \index{gravitational analogy}.

One should stress that this Lorentzian space-time has nothing to do with the real space-time in which our BEC lives. Note also that the invariance of eq. (\ref{kgeq}) under general coordinate transformation is fake. The underlying BEC theory is not even (special) relativistic, but Newtonian, with an absolute time, the laboratory time, with respect to which the equal time commutators (eq. (\ref{etc})) are given.

Said this, one can give a closer look at the metric $g_{\mu\nu}$ given by Eq.(\ref{acm}): a particularly interesting situation is when a steady fluid turns supersonic (i.e. $|\vec v_0|>c$) in some region of space. In a gravitational analogy \index{gravitational analogy}, such a configuration corresponds to a black hole as described in the so-called Painlev\'e -Gullstrand coordinate system and is therefore called a ``sonic black hole": as sound waves travel at a velocity $c$ lower than the fluid velocity $\vec v_0$, they are not able to propagate back and result trapped inside the supersonic region beyond the ``sonic horizon", i.e. the locus where $|\vec v_0|=c$.

In such a setting Eq. (\ref{kgeq}) describes a massless scalar field propagating in a black hole space-time. But this is exactly the system considered by Hawking to obtain his famous result.  One can then repeat step by step Hawking's  derivation  of black hole radiation.
First of all, one has to expand the field  in modes and focus his attention on those upstream propagating modes which are barely able to avoid being trapped by the horizon formation and escape in the subsonic region. Upon quantization, comparison of the  `in' and `out' vacuum states then shows that they are inequivalent since the corresponding annihilation and creation operators are related by a Bogoliubov transformation that mixes them in a non-trivial way. As a result, one can expect that an emission of Bogoliubov phonons by the horizon appears in the sub-sonic region,which are thermally distributed at a temperature given by the surface gravity \index{surface gravity} $\kappa$ of the sonic horizon defined as
\begin{equation}\label{hrsh}
\;\;\;\textrm{with}\;\;\;\kappa=\left.\frac{1}{2c}\frac{d(c^2-\vec v_0^2)}{dn}\right|_{hor} \ ,                                                                                                    \end{equation}
where $n$ is the spatial coordinate normal to the horizon.

It is however crucial to keep in mind that this conclusion is based on a very strong assumption, namely the long wavelength approximation, which has allowed to neglect the last term in eq. (\ref{bodg1}), to rewrite this equation as  $\Box ^2\theta_1=0$, and to introduce the gravitational analogy. As explained in the introduction, the modes of the field  responsible for the Hawking emission  experience an infinite Doppler shift when leaving the near horizon region in the upstream direction. As their wavelength in this region is many order of magnitude smaller than the healing length of the atomic gas, all the derivation of Hawking radiation  in atomic BEC outlined above is at least questionable.

For this reason we go back to the original microscopic BEC theory of Sec.\ref{sec:2} and try to derive Hawking radiation  without making any hydrodynamical  approximation and without any reference to the gravitational analog: the emission of Hawking radiation  in BEC supersonic configurations will then appear as a natural outcome of the underlying  quantum theory.

\section{Stepwise homogeneous condensates}
\label{sec:4}

A simple analytical treatment can be developed to show the occurrence of Bogoliubov phonons creation `` \`a la Hawking" in an atomic BEC undergoing supersonic motion in a very idealized setting consisting of two semi-infinite stationary homogeneous one dimensional condensates (left and right sector) connected by a step-like discontinuity \index{step configuration} \cite{rpc}. As the  purpose of this article is mostly a pedagogical one, we do not enter into a discussion of the actual experimental feasibility of this configuration, for which we refer to the most recent research literature~\cite{larre,kp}.

In particular, we assume the condensate to have a everywhere uniform density $n$ in both sections as well as a spatially uniform flow velocity $v$ along the negative $x$ axis.
The external potential $V_{ext}$ and the repulsive atom-atom interaction coupling $g$
are supposed to be constant within every sector, but to have different values in each sector, satisfying
\begin{equation}\label{cond}
V_{ext}^l+g^l n=V_{ext}^r+g^r n.
\end{equation}
Here, the superscripts ``l" and ``r" refer to left ($x<0$) sector and right ($x>0$) sector respectively, the discontinuity being located at $x=0$. The change in the interaction constant $g$ can be obtained either via the dependence of the atom-atom scattering length on the value of a static external magnetic field, or by modulating the transverse confinement orthogonal to the $x$ direction. Such a change in $g$ directly reflects onto the local sound speed, that then has different values $c^l$ and $c^r$ in the two sections, defined as usual by $m(c^{l,r})^2= n g^{l,r}$.
Thanks to the condition (\ref{cond}), the plane wave form
\begin{equation}
\label{plwa}
\Psi_0(t,x)=\sqrt{n}\,e^{ik_0x-i \omega_0 t}
\end{equation}
of the condensate wavefunction is a solution of the Gross-Pitaevski equation (\ref{gp}) at all times $t$ and positions $x$. The wavevector $k_0$ and the frequency $\omega_0$ are related to the flow velocity $v_0$ by $v_0={\hbar k_0}/{m}$ and $\hbar \omega_0=\hbar^2 k_0^2/ (2m)+gn$.

Let us look now at the solutions of the BdG equation Eq.(\ref{bdg}) for the fluctuation field $\hat \phi$  within each sector. Exploiting the stationarity of the configuration, it is convenient to split $\hat \phi$ into its ``particle'' and ``antiparticle'' components
\begin{equation}\label{frep}
\hat\phi (t,x) =\sum_j \left[ \hat a_j \phi_j (t,x) + \hat a_j^{\dagger} \varphi_j^*(t,x) \right]\ ,\end{equation}
where $\hat a_j$ and $\hat a_j^{\dagger}$ are phonons annihilation and creation operators, satisfying the usual bosonic commutations rules $[\hat a_i,\ \hat a_j^\dagger]=\delta_{ij}$ . The mode functions $\phi_j$ and $\varphi_j$ satisfy the motion equations
\begin{eqnarray}\nonumber
\left[ i(\partial_t + v_0\partial_x) + \frac{\xi c}{2} \partial_x^2 -\frac{c}{\xi} \right] \phi_j &=& \frac{c}{\xi}
\varphi_j\ , \\
\left[ -i (\partial_t + v_0\partial_x) + \frac{\xi c}{2}\partial_x^2 - \frac{c}{\xi}\right] \varphi_j &=&\frac{c}{\xi} \phi_j  \ \label{cde}
\end{eqnarray}
that follow from Eq. (\ref{bdg}) and its conjugate and are to be chosen as oscillating at a frequency $\omega_j$.
Imposing that the equal time commutators satisfy
\begin{equation}\label{etcd}
[\hat \phi (t,x), \hat\phi^{\dagger}(t,x')]=\frac{1}{n}\delta(x-x')\ ,\end{equation}
provides the normalization of the modes
\begin{equation}
\label{nor}
\int dx [\phi_j\phi_{j'}^* - \varphi_j^*\varphi_{j'}]=\pm\frac{\delta_{jj'}}{\hbar n}.
\end{equation}
The sum over $j$ in (\ref{frep}) only involves positive norm modes for which the sign in (\ref{nor}) is positive. 

Within each of the two $x<0$ and $x>0$ spatially uniform regions, the mode functions have a plane wave form
\begin{equation}\label{mdf}
\phi_{\omega}=D(\omega)e^{-i\omega t+ik(\omega)x}\ ,\qquad \varphi_{\omega}=E(\omega)e^{-i\omega t+ik(\omega)x}\ ,
\end{equation}
where  $D(\omega)$  and $E(\omega)$ are normalization factors to be determined using eq. (\ref{nor}). Inserting eqs. (\ref{mdf}) into (\ref{cde}) yields
\begin{eqnarray}\nonumber
\left[ (\omega-v_0k) - \frac{\xi c k^2}{2}  -\frac{c}{\xi} \right] D(\omega) &=& \frac{c}{\xi} E(\omega)\ , \\
\left[ - (\omega-v_0k) - \frac{\xi c k^2}{2} - \frac{c}{\xi}\right] E(\omega) &=& \frac{c}{\xi} D(\omega): \label{gupa}
\end{eqnarray}
the existence of nontrivial solutions requires that the determinant associated to the above homogeneous system vanishes,
\begin{equation}\label{nrela}
(\omega-v_0k)^2=c^2\left(k^2+ \frac{\xi^2 k^4}{4}\right).
\end{equation}
Solving this implicit equation provides the so-called Bogoliubov dispersion \index{Bogoliubov dispersion} of weak excitations on top of a spatially uniform condensate,
\begin{equation}
\omega-v_0k=\pm c\sqrt{k^2+\frac{\xi^2k^4}{4}}\equiv \Omega_{\pm}(k):
\label{eq:bogo_disp}
\end{equation}
here, $\Omega_{\pm}$ is the excitation frequency as measured in the frame co-moving with the fluid. The $+$  $(-)$ sign refers to the positive (negative) norm branch. As expected, for small $k$ such that $k\xi \ll 1$, the dispersion relation \index{dispersion relation} is linear
\begin{equation}
\omega -v_0k =\pm ck\ ,
\label{eq:sonic}
\end{equation}
 this is the hydrodynamical regime to which the gravitational analogy \index{gravitational analogy} is strictly speaking restricted. At higher $k$, the corrections to the linear dispersion are positive and the modes propagate supersonically. For large $k$ such that $k\xi \gg 1$, the relation tends to the quadratic dispersion of single particles.

The normalization condition gives
\begin{equation}\label{nodia}
 |D(\omega)|^2 - |E(\omega)|^2=\pm{1\over 2\pi \hbar n}\Big|\frac{dk}{d\omega}\Big|\
\end{equation}
which using eqs. (\ref{gupa})  yields the normalization factors
\begin{eqnarray}\label{eq:normdispersion}
D(\omega) &=&  \frac{\omega -v_0 k+\frac{c\xi k^2}{2}}{\sqrt{4\pi \hbar n c\xi k^2\left| (\omega-v_0k) \left(\frac{dk}{d\omega}\right)^{-1} \right| }},\nonumber\\
E(\omega) &=& -\frac{\omega -v_0 k-\frac{c\xi k^2}{2}}{\sqrt{4\pi \hbar nc\xi k^2\left| (\omega-v_0k) \left(\frac{dk}{d\omega}\right)^{-1} \right| }}:
\end{eqnarray}
as expected, positive (negative) norm states correspond to the branch of the dispersion relation at a positive (negative) comoving frequency. Remarkably, for any positive norm branch of frequency $\omega$ and wavevector $k$, there exists a negative norm branch of opposite frequency $-\omega$ and wavevector $-k$. Taking advantage of this duality, one can use both positive and negative norm states, replacing the sum over $j$ in (\ref{frep}) with an integral over $\omega$ and restrict to the positive frequency ones.

Let us go back to the dispersion relation  eq. (\ref{nrela}).  At fixed $\omega$ ($>0$) this is a fourth order equation in $k$. It admits four solutions $k^{(i)}_\omega$  and in general $\phi_\omega$  is a linear combination of four plane waves constructed with the $k$'s of the form
\begin{equation}
\phi_\omega(x,t)=e^{-i\omega t}\sum_{i=1}^{4}A_i^{(\omega)}D_i(\omega)e^{ik_\omega^{(i)}x}
\end{equation}
where  the $A_i(\omega)$ are the amplitudes of the modes, not to be confused with the normalization coefficients  $D(\omega)$. Similarly for  $\varphi_\omega(x,t)$.

As said before, our systems consist of two semi-infinite homogeneous condensates joined at x=0 where there is a step-like discontinuity in the speed of sound. Looking at the modes equations eqs (\ref{cde}), one has to require that the solutions in the left region and the ones in the right region satisfy the following matching conditions at $x=0$
\begin{equation}\label{matchingaa}
[\phi]=0,\, [\phi']=0,\, [\varphi]=0,\, [\varphi']=0,
\end{equation}
where $[f(x)]=\lim_{\epsilon\to 0} [f(x+\epsilon)-f(x-\epsilon)]$ and $'$ means $\frac{d}{dx}$.
These four conditions allow to establish a linear relation between the left and right amplitudes
\begin{equation}
A_i^l=M_{ij}A_j^r
\end{equation}
where $M$ is a $4\times 4$ matrix called the matching matrix, not to be confused with the scattering matrix \index{scattering matrix} S we will introduce later, whose dimensionality may vary.

To proceed further in the analysis and explicitly solve the dispersion relation  to get the four roots $k_\omega^{(i)}$, one has to specify the flow configuration under investigation, as the position of the roots in the complex plane varies according to the subsonic or supersonic character of the flow. In the next sections we shall separately consider the different cases.

\section{Subsonic-subsonic configuration}
\label{sec:5}

\begin{figure}[htbp]
\begin{center}
\includegraphics[width=10cm,clip]{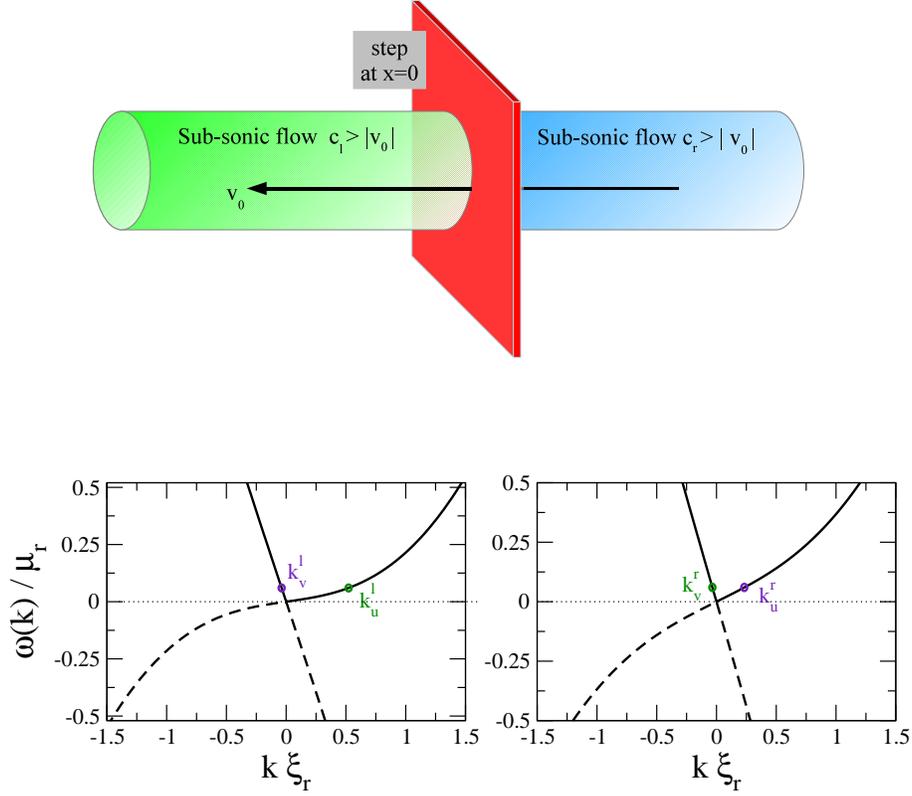}
\includegraphics[width=12cm,clip]{figura_dispersion_subsub.eps}
\end{center}
%
%
\caption{Upper panel: sketch of the subsonic-subsonic flow  configuration. Low panels: dispersion relation of Bogoliubov modes in the asymptotic flat regions away from the transition region. }
\label{fig:subsub_flow}       
\end{figure}

We start by considering a flow which is everywhere sub-sonic \index{subsonic flow}, that is with a flow speed $v_0$ smaller in magnitude than both $c_l$ and $c_r$, that is $|v_0|<c_{r,l}$. A sketch of the configuration under investigation is given in the upper panel of Fig.\ref{fig:subsub_flow}.

\subsection{The Bogoliubov modes and the matching matrix}

The Bogoliubov dispersion in a subsonic flow  is graphically displayed in the two lower panels of Fig.\ref{fig:subsub_flow} for two different relative values of the sound speed $c^l$ (left) and $c^r>c^l$ (right). The positive (negative) norm branches are plotted as solid (dashed) lines.
For any given $\omega>0$, two real solutions belonging to the positive norm branch exist within each $l,r$ region: one, $k_u$, has a positive group velocity $v_g=\frac{d\omega}{dk}$ and propagates in the rightward, upstream direction; the other, $k_v$ has a negative group velocity and propagates in the leftward, downstream direction. The $u,v$ labels used to indicate these solutions are the conventional ones in General Relativity. These two real roots admit a perturbative expansion
\begin{eqnarray}\label{uv}
&&k_v=\frac{\omega}{v_0-c}\left(1+\frac{c^3z^2}{8(v_0-c)^3}+O(z^4)\right)\ ,\nonumber\\
&&k_u=\frac{\omega}{v_0+c}\left(1-\frac{c^3z^2}{8(v_0+c)^3}+O(z^4)\right)\
\end{eqnarray}
where the dimensionless expansion parameter is $z\equiv {\xi \omega/ c}$. To zeroth order in $z$, one recovers the well known hydrodynamical results  $k_v= {\omega}/({v_0 - c})$  and $k_u={\omega}/({v_0 + c})$. In the following, we shall indicate as $k_{u,v}$ the value of these roots in each of the two homogeneous sections on either side of the interface.

The other two solutions of the dispersion relation  are a pair of complex conjugate roots. Within the right sector at $x>0$, we call $k_+^r$  the root with positive imaginary part, which represents a decaying mode when one goes away from the horizon in the positive x direction. The other solution with a negative imaginary part $k_-^r$ corresponds instead to a growing (and therefore non-normalizable) mode.
The opposite holds in the left sector at $x<0$; the $k_+^l$ root with a positive imaginary part represents a growing mode away from the horizon, while the other root $k_-^l$ with a negative imaginary part represents the decaying mode.
Within each $l,r$ region, the wavevector of these modes can be expanded in powers of $z=\xi\omega/c$ as
\begin{multline}\label{decaying}
k_{\pm}=\frac{\omega v_0}{c^2-v_0^2}\left[1-\frac{(c^2+v_0^2)c^4z^2}{4(c^2-v_0^2)^3}+O(z^4)\right]\\
\pm\frac{2i\sqrt{c^2-v_0^2}}{c\xi}\left[1+\frac{(c^2+2v_0^2)c^4z^2}{8(c^2-v_0^2)^3}+O(z^4)\right]\ .
\end{multline}

In summary the decomposition of  $\phi_\omega$  and $\varphi_\omega$ in the left (right) regions reads
\begin{eqnarray}\label{decuno}
\phi_{\omega}^{l(r)} = e^{-i\omega t}\left[A_v^{l(r)}D_v^{l(r)}e^{ik_v^{l(r)}x}\right.&+&A_u^{l(r)}D_u^{l(r)}e^{ik_u^{l(r)}x} + \nonumber \\
&+& \left. A_+^{l(r)}D_{+}^{l(r)}e^{ik_{+}^{l(r)}x}+A_-^{l(r)}D_{-}^{l(r)}e^{ik_{-}^{l(r)}x}\right] \\
 \varphi_{\omega}^{l(r)} = e^{-i\omega t}\left[A_v^{l(r)}E_v^{l(r)}e^{ik_v^{l(r)}x}\right.&+&
  A_u^{l(r)}E_u^{l(r)}e^{ik_u^{l(r)}x}+ \nonumber \\
  &+&\left.A_+^{l(r)}E_{+}^{l(r)}e^{ik_{+}^{l(r)}x}+A_-^{l(r)}E_{-}^{l(r)}e^{ik_{-}^{l(r)}x}\right] \label{decdue}
  \end{eqnarray}
We stress again the fact that the coefficients  $A_{u,v,\pm}^{l(r)}$ are the amplitudes of the different modes, not to be confused with the normalization coefficients, $D_{u,v,\pm}^{l(r)}$ for $\phi_\omega$ and $E_{u,v,\pm}^{l(r)}$ for $\varphi_\omega$: these latter are uniquely fixed  by the commutator relations and the equation of motion, while the amplitudes depend on the choice of basis for the scattering states as we shall see in Sect.~\ref{subsecu:2}. Note that the amplitudes $A_{u,v\pm}^{l(r)}$   are the same for  $\phi_\omega$ and
$\varphi_\omega$ as required by the equation of motion.

The matching conditions at $x=0$, $[\phi] = 0,\ [\varphi]=0,\ [\phi']=0,\ [\varphi']=0$ impose a linear relation between the four left amplitudes $A_{u,v,\pm}^l$ and the right ones $A_{u,v,\pm}^r$
\begin{equation}
W_l\left(
     \begin{array}{c}
       A_v^l \\
       A_u^l \\
       A_+^l \\
       A_-^l \\
 \end{array}
   \right)=W_r\left(
                \begin{array}{c}
                  A_v^r \\
       A_u^r \\
       A_+^r \\
       A_-^r \\
                \end{array}
              \right),
\end{equation}
where the $4\times 4$ matrices $W_{l(r)}$   are
\begin{equation}
\label{eq:wl}
W_{l(r)}=\left(
     \begin{array}{cccc}
       D_v^{l(r)} & D_u^{l(r)} & D_{+}^{l(r)} & D_{-}^{l(r)}\\
       ik_v^{l(r)}D_v^{l(r)} & ik_u^{l(r)}D_u^{l(r)} & ik_+^{l(r)}D_{+}^{l(r)} & ik_-^{l(r)}D_{-}^{l(r)} \\
       E_v^{l(r)} & E_u^{l(r)} & E_{+}^{l(r)} & E_{-}^{l(r)} \\
       ik_v^{l(r)}E_v^{l(r)} & ik_u^{l(r)}E_u^{l(r)} & ik_+^lD_{+}^{l(r)} & ik_-^{l(r)}D_{-}^{l(r)}  \\
\end{array}\right)\,.
\end{equation}
Multiplying both sides by $W_l^{-1}$  one finally gets
\begin{equation}\label{mequa}
     \left( \begin{array}{c}
       A_v^l \\
       A_u^l \\
       A_+^l \\
       A_-^l \\
     \end{array} \right)
   =M \left(
                \begin{array}{c}
                             A_v^r \\
       A_u^r \\
       A_+^r \\
       A_-^r \\
                \end{array}
              \right) \ ,
\end{equation}
in terms of the matching matrix $M=W_l^{-1} W_r$ whose explicit form is rather involved and is not given here.

\subsection{The ``in" and ``out" basis}
\label{subsecu:2}

\begin{figure}[htbp]
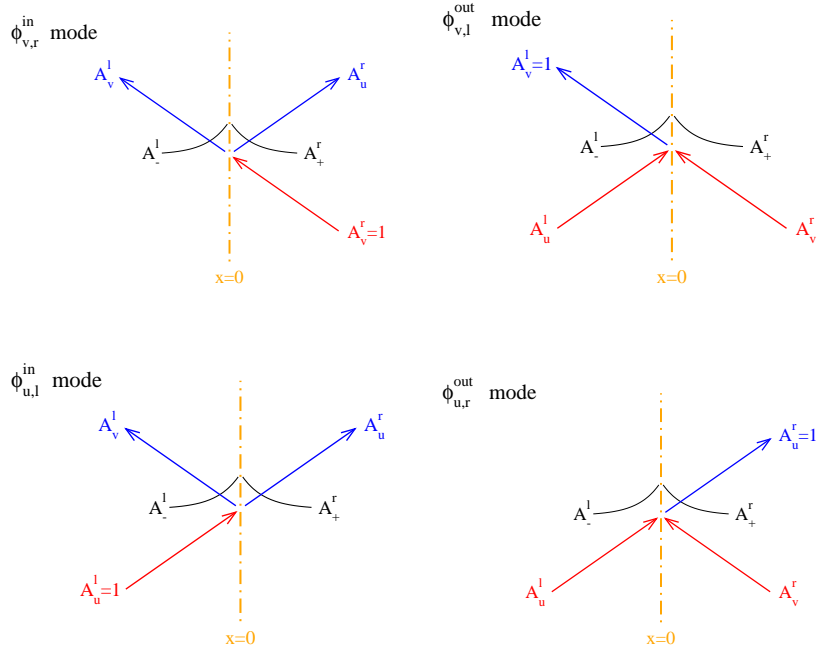

\begin{center}
\includegraphics[width=5.cm,clip]{figura_modes1a.eps}
\hspace{0.5cm}
\includegraphics[width=5.cm,clip]{figura_modes1c.eps} \\
\vspace*{1cm}
\includegraphics[width=5.cm,clip]{figura_modes1b.eps}
\hspace{0.5cm}
\includegraphics[width=5.cm,clip]{figura_modes1d.eps}
\end{center}
%
%
\caption{Sketch of the Bogoliubov modes involved in the ``in'' (left panels) and ``out'' (right panels) basis. The mode labels refer to the dispersion shown in the lower panels of Fig.\ref{fig:subsub_flow}. }
\label{fig:subsub_inout}       
\end{figure}

We now proceed to construct a complete and orthonormal (with respect the scalar product eq. (\ref{nor})) basis for the scattering states of our operator. This can be done in two ways: either choosing a ``in" basis constructed  with incoming scattering states (i.e. states that propagate from the asymptotic regions $x=\pm\infty$ towards the discontinuity at $x=0$) or an ``out" basis constructed with outgoing scattering states (i.e. states that propagate away from the discontinuity to $x=\pm\infty$).

Let us start with the ``in" basis, whose construction is sketched in Fig.\ref{fig:subsub_inout}.
We define the in $v$-mode $\phi_\omega^{v,in}$ as a scattering state with a unit initial amplitude on the left moving $v$ mode incident on the discontinuity from the right ($x=+\infty$), i.e. $D_v^r\ e^{-i\omega t+ik_v^r x}$. The incident wave is scattered by the discontinuity at $x=0$ into a transmitted $v$-mode in the left region with amplitude $A_v^l$ (i.e.   $A_v^l D_v^l\ e^{-i\omega t+ik_v^lx}$) and partially reflected in the right region with amplitude $A_u^r$ (i.e. $A_u^r D_u^r\ e^{-i\omega t+ik_u^rx}$). In order to complete the construction, we have to include in both regions the complex decaying modes as well:  $A_+^r D_+^r e^{-i\omega t+ik_+^rx}$ and $A_-^l D_-^l e^{-i\omega t+ik_-^lx}$. Growing modes are not included as they diverge at infinity.

The general matching equation (\ref{mequa}) becomes in this case
\begin{equation}
\label{modevin}
     \left( \begin{array}{c}
       A_v^l \\
       0 \\
       0\\
       A_-^l\\
     \end{array} \right)
   = M \left(
                \begin{array}{c}
                  1  \\
                  A_u^r \\
                  A_+^r\\
                  0
                \end{array}
              \right) \ .
\end{equation}
Treating $M$ perturbatively in $z_l\equiv \frac{\omega\xi_l}{c_l}$ we obtain (for the simplest case $v_0=0$; for the general subsonic $v_0\neq 0$ case the amplitudes are given in the appendix of \cite{cfr})
\begin{eqnarray}
A_v^l \equiv T&=&\frac{2 \sqrt{c_l c_r}}{c_l+c_r} -\frac{i \sqrt{c_l}  \left(c_l-c_r\right){}^2 z _l}{ c_r^{3/2} \left(c_l+c_r\right)}+\frac{c_l \left(c_l-c_r\right){}^2 \left(c_l^2+c_r^2\right) z _l^2}{2  c_r^3 \left(c_l+c_r\right)^2}\ ,\\
A_u^r \equiv R&=&\frac{c_l-c_r}{c_l+c_r}-\frac{i c_l  \left(c_l-c_r\right){}^2 z _l}{c_r^2 \left(c_l+c_r\right)}-\frac{c_l \left(c_l-c_r\right) \left(2 c_l^3-3 c_l^2 c_r+2 c_l c_r^2+c_r^3\right) z _l^2}{4  c_r^4 \left(c_l+c_r\right)},\ \ \ \ \  \\
A_-^l &=& \frac{\left(c_l-c_r\right) \sqrt{z _l}}{D_-^l \sqrt{c_r} \left(c_l+c_r\right)}-\frac{\left(c_l-c_r\right)z _l^2}{2 D_-^l c_r^{5/2} \left(c_l+c_r\right)}\left[c_r{}^2+i\left(c_l{}^2+c_r{}^2-c_rc_l\right)\right]\ , \\
A_+^r &=& \frac{ c_l \left(-c_l+c_r\right) \sqrt{z _l}}{D_+^r c_r^{3/2} \left(c_l+c_r\right)}+\frac{ c_l^2 \left(c_l-c_r\right)z _l^2}{2D_+^r c_r^{7/2} \left(c_l+c_r\right)}\left[c_l+i\left(c_l-2c_r\right)\right]\ .
\end{eqnarray}
Note that these combine in such a way that the unitarity relation   $|R|^2+|T|^2=1$ is satisfied. Note also that even if they do not enter the unitarity relation, the amplitudes of the decaying modes are part of the full  mode and their presence will show up explicitly contributing to the density-density correlation as we shall see.

In a similar way we can construct the $\phi_\omega^{u,in}$ as a scattering state with a unit initial amplitude on the right moving $u$ mode incident on the discontinuity from the left ($x=-\infty$), which is partially reflected back and partially transmitted, as shown in Fig.\ref{fig:subsub_inout}. Here too, we have to include the decaying modes.

The matching relation now reads
\begin{equation}
\label{modeuin} 
     \left( \begin{array}{c}
       A_v^l \\
       1  \\
       0\\
       A_-^l\\
     \end{array} \right)
   = M\left(
                \begin{array}{c}
                  0  \\
                  A_u^r \\
                  A_+^r\\
                  0
                \end{array}
              \right) \
\end{equation}
yielding
\begin{eqnarray}
  A_v^l \equiv R'&=& \frac{c_r-c_l}{c_l+c_r}-\frac{i   \left(c_l-c_r\right){}^2 z _l}{c_r \left(c_l+c_r\right)}+\frac{\left(c_l-c_r\right) \left(c_l^3+2 c_l^2 c_r-3 c_l c_r^2+2 c_r^3\right) z _l^2}{4  c_r^3 \left(c_l+c_r\right)}\ \\
A_u^r\equiv T' &=&\frac{2 \sqrt{c_l c_r}}{c_l+c_r}-i\frac{ \sqrt{c_l}  \left(c_l-c_r\right){}^2 z _l}{ c_r^{3/2} \left(c_l+c_r\right)}-\frac{\sqrt{c_l} \left(c_l-c_r\right){}^2 \left(c_l^2-4 c_l c_r+c_r^2\right) z _l^2}{8  c_r^{7/2} \left(c_l+c_r\right)}\ \\
A_-^l&=&\frac{  \left(c_l-c_r\right) \sqrt{z_l}}{D_-^l \sqrt{c_l} \left(c_l+c_r\right)}+\frac{  \left(c_l-c_r\right)}{2D_-^l\sqrt{c_l}  c_r \left(c_l+c_r\right)}\left[-c_r+i\left(2c_l-c_r\right)\right]z_l^2\ \\
  A_+^r&=&\frac{  \sqrt{c_l} \left(-c_l+c_r\right) \sqrt{z _l}}{D_+^r c_r \left(c_l+c_r\right)}+\frac{\sqrt{c_l}\left(c_l-c_r\right)}{2 D_+^r  c_r^3 \left(c_l+c_r\right)}\left[c_l{}^2+i\left(c_l{}^2+c_r{}^2-c_lc_r\right)\right]z _l^2\ \ \ \
\end{eqnarray}
which implies    $|R'|^2+|T'|^2=1$, as required by unitarity.

The scattering modes  $\phi_\omega^{v,in}$ and $\phi_\omega^{u,in}$, and the similarly constructed  $\varphi_\omega^{v,in}$ and $\varphi_\omega^{u,in}$, constitute a complete ``in" basis for our field operator $\hat \phi$, that can be then expanded as
\begin{equation}
\hat\phi(x,t)=\int_{0}^{\infty}d\omega\Big[\hat a_{\omega}^{v,in}\phi_{v,r}^{in}(t,x)
+\hat a_{\omega}^{u,in}\phi_{u,l}^{in}(t,x)
+\hat a_{\omega}^{v,in\dagger}\varphi_{v,r}^{in*}(t,x)
+\hat a_{\omega}^{u,in\dagger}\varphi_{u,l}^{in*}(t,x)\Big].
\end{equation}
The ``in" vacuum  $|0,in\rangle$ is defined as usual by $\hat a_\omega^{u,in}|0,in\rangle = 0$ and $\hat a_\omega^{v,in}|0,in\rangle = 0$. The $N$-phonons states that constitute the ``in" basis of the Hilbert space are constructed by a repeated action of the creation operators $\hat a_\omega^{\dagger u,in}$ and $\hat a_\omega^{\dagger v,in}$ on the vacuum state.

While the ``in" basis has been constructed using incoming scattering modes, an alternative ``out" basis can be constructed starting from the outgoing scattering basis of the $\hat \phi$ field operator, composed of modes that emerge from the scattering region around $x=0$ with unit amplitude on a wave propagating at $t=+\infty$ either rightwards towards $x=+\infty$ or leftwards towards $x=-\infty$.

We begin by defining the $\phi_\omega^{v,out}$ scattering mode: as it is sketched in Fig.~\ref{fig:subsub_inout},  this is a linear combination of in-going right and left moving components with amplitudes $A_u^l$ and $A_v^r$ and decaying modes with amplitudes $A_-^l$ and $A_+^r$. These coefficients are chosen in a way to give after scattering only a left moving $v$-mode of unit amplitude. This imposes the condition:
\begin{equation}
\label{modevout}
     \left( \begin{array}{c}
       1 \\
       A_u^l  \\
       0\\
       A_-^l\\
     \end{array} \right)
   = M\left(
                \begin{array}{c}
                  A_v^r  \\
                  0 \\
                  A_+^r\\
                  0
                \end{array}
              \right) \
\end{equation}
that yields
\begin{eqnarray}
  A_u^l \equiv R'^{*}&=&\frac{c_r-c_l}{c_l+c_r}+\frac{i \left(c_l-c_r\right){}^2 z _l}{ c_r \left(c_l+c_r\right)}+\frac{ \left(c_l-c_r\right) \left(c_l^3+2 c_l^2 c_r-3 c_l c_r^2+2 c_r^3\right) z _l^2}{4  c_r^3 \left(c_l+c_r\right)}\ ,\\
A_v^r \equiv T'^{*}&=&\frac{2 \sqrt{c_l c_r}}{c_l+c_r} +\frac{i \sqrt{c_l}  \left(c_l-c_r\right){}^2 z _l}{ c_r^{3/2} \left(c_l+c_r\right)}-\frac{ \left(c_l-c_r\right){}^2 \left(c_l^2-4 c_l c_r+c_r^2\right) z _l^2}{8 c_r^{7/2} \left(c_l+c_r\right)}\ ,\\
A_-^l&=&\frac{ \left(c_l-c_r\right) \sqrt{z _l}}{D_-^l \sqrt{c_l}  \left(c_l+c_r\right)}-\frac{ \left(c_l-c_r\right)z _l^2}{2D_-^l  c_r \left(c_l+c_r\right)}\left[c_r+i\left(2c_l-c_r\right)\right]\ ,\\
  A_+^r&=&\frac{\sqrt{c_l} \left(-c_l+c_r\right) \sqrt{z _l}}{D_+^r c_r \left(c_l+c_r\right)}+\frac{\sqrt{c_l} \left(c_l-c_r\right)z _l^2}{2 D_+^r  c_r^3 \left(c_l+c_r\right)}\left[c_l{}^2-i\left(c_l{}^2+c_r{}^2-c_rc_l\right)\right]\ .
\end{eqnarray}
with $|R'^*|^2 + |T'^*|^2=1$.

The same procedure can be used to construct the mode $\phi_\omega^{u,out}$, by imposing the out-going waves to consist of a unit amplitude right moving $u$-mode only.
In this case, the matching relations are
\begin{equation}
\label{modeuout}
     \left( \begin{array}{c}
       0 \\
       A_u^l  \\
       0\\
       A_-^l\\
     \end{array} \right)
   = M\left(
                \begin{array}{c}
                  A_v^r  \\
                  1 \\
                  A_+^r\\
                  0
                \end{array}
              \right) \
\end{equation}
with
\begin{eqnarray}
A_u^l \equiv T^* &=&\frac{2 \sqrt{c_l c_r}}{c_l+c_r} +\frac{i \sqrt{c_l}  (c_l-c_r)^2 z_l}{c_r^{3/2} \left(c_l+c_r\right)}-\frac{\sqrt{c_l} \left(c_l-c_r\right){}^2 \left(c_l^2-4 c_l c_r+c_r^2\right) z_l^2}{8  c_r^{7/2} \left(c_l+c_r\right)}\ ,\\
A_v^r \equiv R^* &=&\frac{c_l-c_r}{c_l+c_r}+\frac{i c_l \left(c_l-c_r\right){}^2 z_l}{c_r^2 \left(c_l+c_r\right)}-\frac{c_l\left(c_l-c_r\right) \left(2 c_l^3-3 c_l^2 c_r+2 c_l c_r^2+c_r^3\right) z_l^2}{4 c_r^4 \left(c_l+c_r\right)}\ \ \ \ \ \ \\
A_-^l&=&\frac{c_l \left(c_l-c_r\right) z_l}{D_-^l \sqrt{c_r} \left(c_l+c_r\right)}+\frac{\left(c_l-c_r\right)z _l^2}{2 D_-^l  c_r^{5/2} \left(c_l+c_r\right)}\left[-c_r{}^2+i\left(c_l{}^2+c_r{}^2-c_lc_r\right)\right], \ \ \ \ \\
A_+^r&=&\frac{c_l \left(-c_l+c_r\right) z_l}{D_+^r c_r^{3/2} \left(c_l+c_r\right)}+\frac{  c_l^2 \left(c_l-c_r\right)z_l^2}{2D_+^r c_r^{7/2} \left(c_l+c_r\right)}\left[c_l+i\left(2c_r-c_l\right)\right]\ .
\end{eqnarray}
In analogy to what was done for the ``in" basis, this ``out" basis can be used to obtain a decomposition of the $\hat \phi$ field operator as
\begin{equation}
\hat\phi(x,t)=\int_{0}^{\infty}d\omega\Big[\hat a_{\omega}^{v,out}\phi_{v,l}^{out}(t,x)
+\hat a_{\omega}^{u,out}\phi_{u,r}^{out}(t,x)
+\hat a_{\omega}^{v,out\dagger}\varphi_{v,l}^{out*}(t,x)
+\hat a_{\omega}^{u,out\dagger}\varphi_{u,r}^{out*}(t,x)\Big]
\end{equation}
in terms of the bosonic annihilation and creation operators for the out-going modes.
This also leads to an alternative vacuum state defined by the conditions
$\hat a_\omega^{u,out}|0,out\rangle = \hat a_\omega^{v,out}|0,out\rangle = 0$ and an alternative ``out" basis of the Hilbert space.

\subsection{Bogoliubov transformation}

As both the ``in" and the ``out" basis are complete,  the ``in" and ``out" scattering modes can be related by the simple linear scattering relations
\begin{eqnarray}\label{rt}
  \phi_{v,r}^{in} &=& T \phi_{v,l}^{out}+R \phi_{u,r}^{out}, \nonumber\\
  \phi_{u,l}^{in} &=& R' \phi_{v,l}^{out}+T' \phi_{u,r}^{out},
\end{eqnarray}
that can be summarized in terms of a unitary $2\times 2$ scattering matrix $S$
\begin{equation}\label{smatsub}
S=\left(
  \begin{array}{cccc}
   T & R \\
     R' & T' \\
  \end{array}\right).
\end{equation}
Analogous relations hold for the $\varphi_\omega$ modes.

Expressed in terms of mode amplitudes, these scattering relations define a linear Bogoliubov transformation \index{Bogoliubov transformation} relating the annihilation and creation operators for the ``out" modes to the ones of the ``in" modes. In the specific case of the sub-sub interface considered in the present section, all $u$ and $v$ modes involved in the scattering process have positive norm, so there is no mixing of the annihilation and creation operators:
\begin{eqnarray}
  \hat a_{\omega}^{v,out} &=& T\hat a_{\omega}^{v,in}+R'\hat a_{\omega}^{u,in}, \nonumber\\
  \hat a_{\omega}^{u,out} &=& R\hat a_{\omega}^{v,in}+T'\hat a_{\omega}^{u,in}.
\end{eqnarray}
As a result, the Bogoliubov transformation trivially reduces to a unitary transformation of the ``in" and ``out" Hilbert space that conserves the number of excitations and, in particular, preserves the vacuum state: if the system is initially in the  $|0,in\rangle$ state with no  incoming particles, the number of outgoing particles will also be zero,
\begin{multline}
n_\omega^{v(u),out}= \langle 0,in|\hat a_\omega^{v(u),out\dagger}a_\omega^{ v(u),out}|0,in\rangle = \\
=\langle 0,in|
  (T^*(R^*) \hat a_{\omega}^{v,in\dagger}+R'^*(T'^*)\hat a_{\omega}^{u,in\dagger})
(T(R)\hat a_{\omega}^{v,in}+R'(T') \hat a_{\omega}^{u,in})
|0,in\rangle=0.
\end{multline}
No phonon can be created, but only scattered at the horizon.

\subsection{Density-density correlations \index{density-density correlations}}

Correlation functions \index{correlation functions} are a modern powerful tool to investigate the properties of strongly correlated atomic gases~\cite{correl}.
We shall concentrate our attention to the correlation pattern of the density fluctuations at equal time, defined as
\begin{equation}\label{densitya}
G^{(2)}(t;x,x')\equiv \frac{1}{2n^2}\lim_{t\rightarrow t'}\langle {\rm in}|\{ \hat n_1 (t,x), \hat n_1 (t',x')\}|{\rm in}\rangle\ \ ,
\end{equation}
where $\{,\}$ denotes the anticommutator. In our configuration with a spatially uniform condensate density $n$, the density fluctuation operator $\hat n_1$ can be expanded in the in-going annihilation and creation operators as
\begin{multline}
\hat n^1\equiv n(\hat \phi(x,t) + \hat \phi^{\dagger}(x,t))=\\
=n\int_0^{\infty}d\omega\left[\hat a_{\omega}^{v,in}(\phi_{v,r}^{in}+\varphi_{v,r}^{in})+
\hat a_{\omega}^{u,in}(\phi_{u,l}^{in}+\varphi_{u,l}^{in})+{\rm h.c.}\right].
\end{multline}
or, alternatively, one can use the ``out" basis.

In either case, by evaluating  $G^{(2)}$ on the vacuum state $|0,in\rangle=|0,out\rangle$, one finds for one point located to the left ($x<0$) and the other to the right ($x>0$) the not too significative expression
\begin{multline}
G^{(2)}(t;x,x')\simeq -\frac{\hbar}{2\pi mn(c_r+c_l)}\left[\frac{1}{(v_0-c_l)(v_0-c_r)\left(\frac{x}{c_l-v_0} +\frac{x'}{v_0-c_r}\right)^2}\right.+ \\
+\left.
\frac{1}{(v_0+c_l)(v_0+c_r)\left(-\frac{x}{v_0+c_l}+\frac{x'}{v_0+c_r}\right)^2}\right],
\end{multline}
which just shows correlations decreasing with the square of the distance weighted by the effective speed of sound in the different regions. The physical origin of these correlations is traced back to the repulsive interactions between particles in the gas.

\section{Subsonic-supersonic configuration}
\label{sec:6}

\subsection{the modes and the matching matrix}

\begin{figure}[htbp]
\begin{center}
\includegraphics[width=10cm,clip]{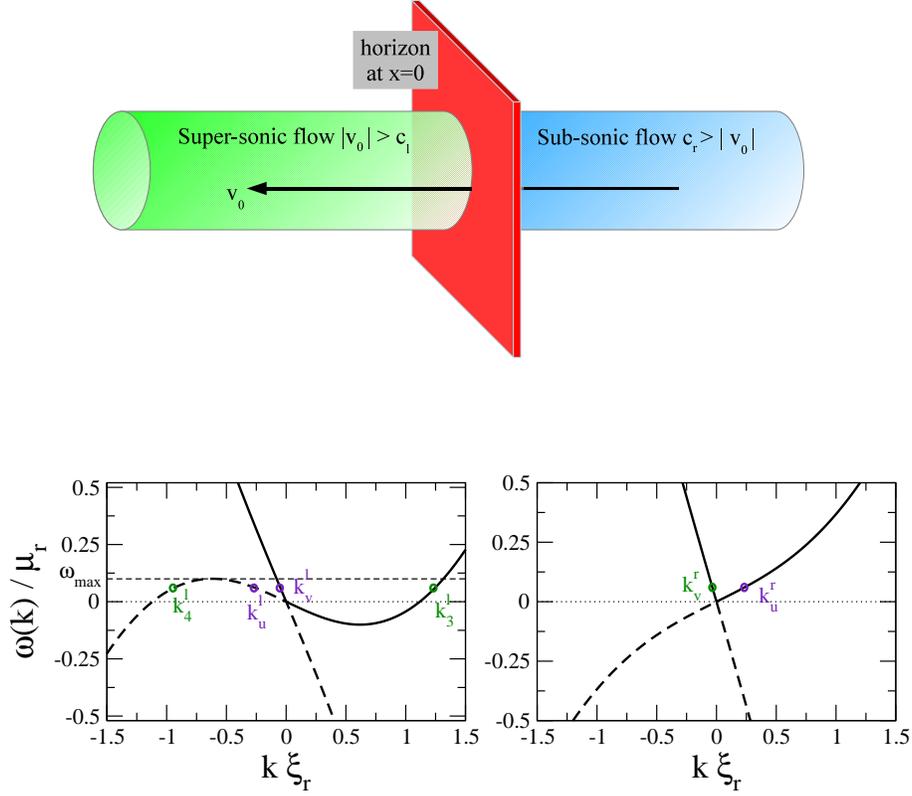}
\includegraphics[width=12cm,clip]{figura_dispersion_subsuper.eps}
\end{center}
%
%
\caption{Upper panel: sketch of the subsonic-supersonic flow configuration. Low panels: dispersion relation \index{dispersion relation} of Bogoliubov modes in the asymptotic regions away from the horizon. }
\label{fig:subsuper_flow}       
\end{figure}

The warm up exercise discussed in detail in the previous section has allowed to take confidence with the formalism. In this section, we shall consider the much more interesting case of the {\em acoustic black hole \index{acoustic black hole}} configuration sketched in Fig.\ref{fig:subsuper_flow}: taking again the flow to be in the negative $x$ direction ($v_0<0$), we assume that the flow is sub-sonic $c_r>|v_0|$ in the upstream $x>0$ sector, while it is super-sonic $c_l<|v_0|$ in the downstream $x<0$ sector \index{supersonic flow}.

The analogy with a gravitational black hole is simply understood: long wavelength sound waves in the $x<0$ supersonic region are dragged away by the flow and no longer able to propagate in the upstream direction. The outer boundary of the super-sonic region separating it from the sub-sonic one plays the role of the horizon: long wavelength sound waves can cross it only in the direction of the flow, and eventually get trapped inside the acoustic black hole . Even if this picture perfectly captures the dynamics of long wavelength  Bogoliubov waves in the sonic window where the dispersion has the form Eq. (\ref{eq:sonic}), the supersonic correction that is visible in Eq.(\ref{eq:bogo_disp}) introduces remarkable new effetcs as we shall see in the following of the section.

The analysis of the dispersion relation  and of the modes in the $x>0$ subsonic region on the right of the horizon is the same as given in the previous section: two oscillating modes exist with real wave vectors $k_{u(v)}^r$ as well as two complex conjugate evanescent modes with $k_{\pm}^r$.

In the $x<0$ supersonic region on the left of the horizon, the dispersion relation  has a significantly different shape, as shown in the lower-left panel Fig.\ref{fig:subsuper_flow}.
In particular, it is immediate to see that there exists a threshold frequency $\omega_{max}$ above which the situation resembles the one of the subsonic regime: two oscillatory modes exist propagating in the downstream and upstream directions, respectively. Note that the upstream propagation occurs in spite of the super-sonic character of the underlying flow because of the super-luminal dispersion of Bogoliubov waves predicted by Eq.(\ref{eq:bogo_disp}). Of course, this mode falls well outside the sonic region where the hydrodynamic approximation is valid. The threshold frequency $\omega_{max}$ is given by the maximum frequency of the negative norm Bogoliubov mode as indicated in the lower-left panel Fig.\ref{fig:subsuper_flow}. In formulas, it corresponds to the Bogoliubov frequency of the mode at a $k_{max}$ value such that
\begin{equation}
k_{max}=-\frac{1}{\xi_l}\,\left[-2+\frac{v_0^2}{2c_l^2}+\frac{|v_0|}{2c_l}\sqrt{8+\frac{v_0^2}{c_l^2}}\right]^{1/2}.
\end{equation}

The $0<\omega<\omega_{max}$ case is much more interesting: from Fig.~\ref{fig:subsuper_flow}, one sees that four real roots of the dispersion relation  exist, corresponding to four oscillatory modes, two on the positive norm branch and two on the negative norm one. Two of these modes denoted as $u,v$ lie in the small $k$ region at
\begin{eqnarray}
k_v&=&\frac{\omega}{v-c_l}\left[1+\frac{c_l^3z_l^2}{8(v_0-c_l)^3}+O(z_l^2)\right] \\
k_u&=&\frac{\omega}{v+c_l}\left[1-\frac{c_l^3z_l^2}{8(v_0+c_l)^3}+O(z_l^2)\right]
\end{eqnarray}
and have a hydrodynamic character. Differently from the sub-sonic case, both of them propagate in the downstream direction with a negative group velocity $\frac{d\omega}{dk}<0$: also the $u$ mode that in the comoving frame with the fluid propagates to the right is dragged by the super-sonic flow and turns out to be forced to propagate in the left direction.
Furthermore, while the $k_v$ solution belongs as before to the positive norm branch, the $k_u$ solution belongs now to the negative norm branch and the corresponding excitation quanta carry a negative energy $\omega<0$.

The wavevector of the other two roots indicated as $k_3$ and $k_4$ in the figure is non-perturbative in $\xi$
\begin{multline}
k_{3,4}=\frac{\omega v_0}{c_l^2-v_0^2}\left[1-\frac{(c_l^2+v_0^2)c_l^4z_l^2}{4(c_l^2-v_0^2)^3}+O(z_l^4)\right]+ \\
\pm\frac{2\sqrt{v_0^2-c_l^2}}{c_l\xi_l}\left[1+\frac{(c_l^2+2v_0^2)c_l^4z_l^2}{8(c_l^2-v_0^2)^3}+O(z_l^4)\right]\ .
\end{multline}
and lies well outside the hydrodynamic region. Comparing these roots with Eq.(\ref{decaying}), one realizes that $k_{3,4}$ are the analytic continuation for supersonic flow \index{supersonic flow} of the growing and decaying modes previously discussed for the subsonic regime. The $k_3$ mode belongs to the positive norm branch, while $k_4$ to the negative one;  both of them have a positive group velocity and propagate in the upstream direction.

For $\omega<\omega_{max}$, the general solution of the modes equation in the super-sonic  (left) region reads then
\begin{eqnarray}
  \phi_{\omega}^{l} &=& e^{-i\omega t}\left[A_v^{l}D_v^{l}e^{ik_v^{l}x}+A_u^{l}D_u^{l}e^{ik_u^{l}x}+A_3^{l}D_{3}^{l}e^{ik_3^{l}x}
+A_4^{l}D_{4}^{l}e^{ik_4^{l}x}\right], \nonumber\\
  \varphi_{\omega}^{l} &=& e^{-i\omega t}\left[A_v^{l}E_v^{l}e^{ik_v^{l}x}+
  A_u^{l}E_u^{l}e^{ik_u^{l}x}+A_3^{l}E_{3}^{l}e^{ik_3^{l}x}+A_4^{l}E_{4}^{l}e^{ik_4^{l}x}\right] \nonumber\ ,
  \end{eqnarray}
while in the sub-sonic (right) region it reads
\begin{equation}
  \phi_{\omega}^{r} = e^{-i\omega t}\left[A_v^{r}D_v^{r}e^{ik_v^{r}x}+A_u^{r}D_u^{r}e^{ik_u^{r}x}+A_+^{r}D_{+}^re^{ik_+^{r}x}
+A_-^{r}D_{-}^r e^{ik_-^{r}x}\right].
\end{equation}
An analogous expression holds for $\varphi_\omega^r$ once we replace $D(\omega)$ with $E(\omega)$.

As we have discussed in the sub-sub case, the field amplitudes on the left and the right of the discontinuity point at $x=0$ are related by
\begin{equation}
\label{eq:matchingtc}
     \left( \begin{array}{c}
       A_v^l \\
       A_u^l \\
       A_3^l \\
       A_4^l \\
     \end{array} \right)
   =M \left(
                \begin{array}{c}
                             A_v^r \\
       A_u^r \\
       A_+^r \\
       A_-^r \\
                \end{array}
              \right) \ ,
\end{equation}
the matching matrix being written as $M =W_l^{-1}W_r$ in terms of $W_r$ given by Eq. (\ref{eq:wl}) and
\begin{equation}
\label{eq:wla}
W_{l}=\left(
     \begin{array}{cccc}
       D_v^{l} & D_u^{l} & D_{3}^{l} & D_{4}^{l}\\
       ik_v^{l}D_v^{l} & ik_u^{l}D_u^{l} & ik_3^{l}D_{3}^{l} & ik_4^{l}D_{4}^{l} \\
       E_v^{l} & E_u^{l} & E_{3}^{l} & E_{4}^{l} \\
       ik_v^{l}E_v^{l} & ik_u^{l}E_u^{l} & ik_3^lD_{3}^{l} & ik_4^{l}D_{4}^{l}  \\
\end{array}\right).
\end{equation}

\subsection{The ``in" and ``out" basis}

\begin{figure}[htbp]
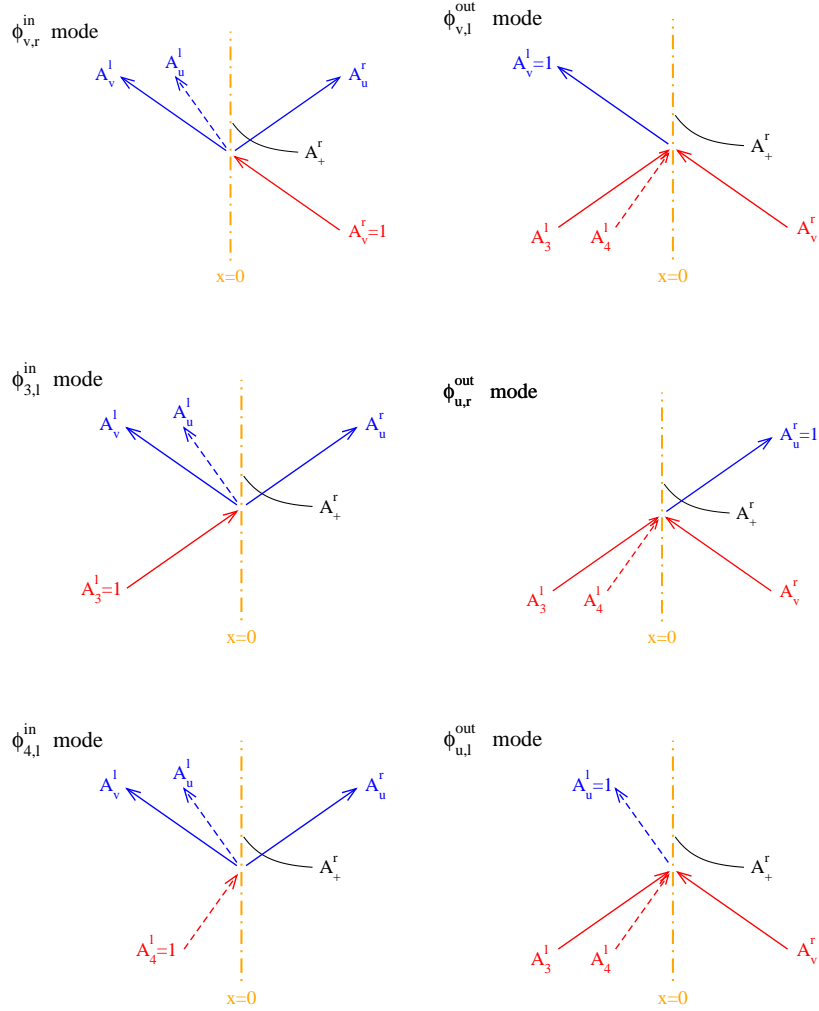

\begin{center}
\includegraphics[width=5.cm,clip]{figura_modes2a.eps}
\hspace{0.5cm}
\includegraphics[width=5.cm,clip]{figura_modes2d.eps} \\
\vspace*{1cm}
\includegraphics[width=5.cm,clip]{figura_modes2b.eps}
\hspace{0.5cm}
\includegraphics[width=5.cm,clip]{figura_modes2e.eps}\\
\vspace*{1cm}
\includegraphics[width=5.cm,clip]{figura_modes2c.eps}
\hspace{0.5cm}
\includegraphics[width=5.cm,clip]{figura_modes2f.eps}
\end{center}
%
%
\caption{Sketch of the Bogoliubov modes involved in the ``in'' (left panels) and ``out'' (right panels) basis. The mode labels refer to the dispersion shown in the lower panels of Fig.\ref{fig:subsuper_flow}. }
\label{fig:subsuper_inout}       
\end{figure}

We can now proceed to construct the ``in" scattering basis. Differently from the sub-sub case discussed in the previous section, there are now three ``in" scattering modes associated to the processes sketched in Fig.\ref{fig:subsuper_inout}.

The mode $\phi_\omega^{v,in}$ is defined as an initial left-moving unit amplitude $v$ wave propagating in the $x>0$ sub-sonic region towards the horizon, which upon scattering generates in the subsonic region a reflected right-moving $u$ wave of amplitude $A_u^r$  and a spatially decaying wave of amplitude $A_+^r$. The transmitted waves in the $x<0$ supersonic region are now in the number of two, and both travel in the leftward direction along the flow. One is the standard transmitted $v$ wave, with positive norm and amplitude $A_v^l$, the other is a negative norm $u$ wave with amplitude $A_u^l$, the so-called {\em anomalous transmitted} wave.

To leading order in $z_l$, the corresponding amplitudes are
\begin{eqnarray}\label{vin}
  A_v^l &=& \sqrt{\frac{c_r}{c_l}}\frac{v_0-c_l}{v_0-c_r}
=S_{vl,vr}, \label{eq:Svl,vr} \\
  A_u^r &=& \frac{v_0+c_r}{v_0-c_r} =S_{ur,vr}, \\
  A_u^l &=&  \sqrt{\frac{c_r}{c_l}}\frac{v_0+c_l}{c_r-v_0} =S_{ul,vr}, \\
  A_+^r &=& \frac{c_l\sqrt{z_l}\sqrt{c_r(v_0^2-c_l^2)}}{\sqrt{2}D_{+}^r(v_0-c_l)(c_r^2-v_0^2)^{3/2}(c_r+c_l)}\Big[\sqrt{c_r^2-v_0^2}\left(v+\sqrt{v_0^2-c_l^2}\right)+
\nonumber \\
&+&i\left(v_0\sqrt{v_0^2-c_l^2}+v_0^2-c_r^2\right)\Big]=S_{+r,vr}. \label{eq:S+r,vr}
\end{eqnarray}
Note the shorthand notation introduced in (\ref{eq:Svl,vr}-\ref{eq:S+r,vr}) to simply identify the incoming and outgoing channel: for example the matrix element $S_{ul,vr}$ indicates that the incoming channel (second index)  is a $v$-mode entering from the right region, while the outgoing channel (first index) is a $u$-mode escaping in the left region.
The conservation of the Bogoliubov norm translates into a unitary condition between the amplitudes of the propagating modes,
\begin{equation}
|A_v^l|^2+|A_u^r|^2-|A_u^l|^2=1,
\end{equation}
where the minus sign comes from the negative norm $u,l$-mode.

As it is sketched in Fig.\ref{fig:subsuper_inout}, the other two ``in" scattering modes $\phi_\omega^{3,in}$ and $\phi_\omega^{4,in}$ are constructed in a similar way by imposing a unit amplitude in the $k_3$ or $k_4$ waves incident on the horizon from the left supersonic side. For the $\phi_\omega^{3,in}$ ``in" scattering mode, the corresponding amplitudes are given by
\begin{eqnarray}\label{3in}
  A_v^l &=& \frac{(v_0^2-c_l^2)^{3/4}(v_0+c_r)}{c_l^{3/2}\sqrt{2z_l}(c_l+c_r)\sqrt{c_r^2-v_0^2}}\left(\sqrt{c_r^2-v_0^2}+i\sqrt{v_0^2-c_l^2}\right)
=S_{vl,3l},  \\
  A_u^r &=& \frac{\sqrt{2c_r}(v_0^2-c_l^2)^{3/4}(v_0+c_r)}{c_l\sqrt{z_l}(c_r^2-c_l^2)\sqrt{c_r^2-v_0^2}}\left(\sqrt{c_r^2-v_0^2}+i\sqrt{v_0^2-c_l^2}\right)
=S_{ur,3l}, \\
  A_u^l &=& \frac{(v_0^2-c_l^2)^{3/4}(v_0+c_r)}{c_l^{3/2}\sqrt{2z_l}(c_l-c_r)\sqrt{c_r^2-v_0^2}}
\left(\sqrt{c_r^2-v_0^2}+i\sqrt{v_0^2-c_l^2}\right)=S_{ul,3l}, \\
  A_+^r &=& \frac{(v_0^2-c_l^2)^{1/4}}{2D_{+}^r(v_0^2-c_r^2)}(v_0-i\sqrt{c_r^2-v_0^2})=S_{+r,3l}.
\end{eqnarray}
and satisfy the unitarity relation
\begin{equation}
|A_v^l|^2+|A_u^r|^2-|A_u^l|^2=1.
\end{equation}
As a most remarkable point, note that the amplitudes for the propagating modes now diverge at small $\omega$ as $\frac{1}{\sqrt{\omega}}$.

For the $\phi_\omega^{4,in }$ ``in" scattering mode one gets instead
\begin{eqnarray}\label{4in}
  A_v^{l} &=& \frac{(v_0^2-c_l^2)^{3/4}(v_0+c_r)}{c_l^{3/2}\sqrt{2z_l}(c_l+c_r)\sqrt{c_r^2-v_0^2}}\left(\sqrt{c_r^2-v_0^2}-i\sqrt{v_0^2-c_l^2}\right)\ =S_{vl,4l}\ \ , \\
  A_u^{r} &=& \frac{\sqrt{2c_r}(v_0^2-c_l^2)^{3/4}(v_0+c_r)}{c_l\sqrt{z_l}(c_r^2-c_l^2)\sqrt{c_r^2-v_0^2}}\left(\sqrt{c_r^2-v_0^2}-i\sqrt{v_0^2-c_l^2}\right) =S_{ur,4l}\ \ ,\\
  A_u^{l} &=& \frac{(v_0^2-c_l^2)^{3/4}(v_0+c_r)}{c_l^{3/2}\sqrt{2z_l}(c_l-c_r)\sqrt{c_r^2-v_0^2}}\left(\sqrt{c_r^2-v_0^2}-i\sqrt{v_0^2-c_l^2}\right)\ =S_{ul,4l}\ \ ,\\
  A_+^{r} &=& \frac{(v_0^2-c_l^2)^{1/4}(v_0^2-c_l^2+v_0\sqrt{v_0^2-c_l^2})}{2D_{+}(c_r^2-v_0^2)(c_l^2-v_0^2+v_0\sqrt{v_0^2-c_l^2})}(v_0-i\sqrt{c_r^2-v_0^2})\ =S_{+r,4l}\ \ .
\end{eqnarray}
and the unitarity condition reads
\begin{equation}
|A_v^l|^2+|A_u^r|^2-|A_u^l|^2=-1:
\end{equation}
the minus sign on the right-hand side comes from the fact that the incoming unit amplitude $k_4$ mode has negative norm. All together, the $v$, $3$ and $4$ ``in" scattering modes form a basis on which to expand the $\hat \phi$  field operator.

The construction of the ``out" basis proceeds along similar lines: one has the three $\phi_{v,l}^{out}$, $\phi_{u,r}^{out}$ and $\phi_{u,l}^{out}$ ``out" scattering modes, where the $l(r)$ label near the superscript $u$ indicates again the left (right) region of space. The corresponding scattering processes are depicted in Fig.\ref{fig:subsuper_inout}.

As the corresponding amplitudes will not ne needed in the following, we refer the reader to Ref.\cite{cfr} for their explicit expression. As discussed in \cite{machpare,rpc}, the field operator can then be equivalently expanded in the basis of the ``in" scattering modes as
\begin{multline}
\hat\phi=\int_{0}^{\omega_{max}}d\omega\left[\hat a_{\omega}^{v,in}\phi_{v,r}^{in}+\hat a_{\omega}^{3,in}\phi_{3,l}^{in}+\hat a_{\omega}^{4,in\dagger}\phi_{4,l}^{in}+\right.
 \\ \left.+ \hat a_{\omega}^{v,in\dagger}\varphi_{v,r}^{in*}+\hat a_{\omega}^{3,in\dagger}\varphi_{3,l}^{in*}+\hat a_{\omega}^{4,in}\varphi_{4,l}^{in*}\right],
 \end{multline}
or, equivalently on the basis of the ``out" scattering ones. Note in particular the third term on the right-hand side, as the corresponding $k_4$ mode is a negative norm one, this term enters with a $\phi_{4,l}^{in }$ field multiplied by a creation $\hat a^{4,in\dagger}_\omega$ operator: as we shall see in the next sub-section, this simple fact is the key element leading to the emission of analog Hawking radiation \index{Hawking radiation} by the horizon.

\subsection{Bogoliubov transformation}

The ``in" and ``out" basis are now related by a $3\times 3$ scattering matrix $S$ relating the three incoming states to the three outgoing states. Explicitly
\begin{eqnarray}\label{eq:outinaa}
  \phi_{v,r}^{in}&=& S_{vl,vr} \phi_{v,l}^{out}+S_{ur,vr} \phi_{u,r}^{out}+S_{ul,vr} \phi_{u,l}^{out}\ , \\
  \phi_{3,l}^{in} &=& S_{vl,3l} \phi_{v,l}^{out}+S_{ur,3l} \phi_{u,r}^{out}+S_{ul,3l} \phi_{u,l}^{out}\ , \\
  \phi_{4,l}^{in} &=& S_{vl,4l} \phi_{v,l}^{out}+S_{ur,4l} \phi_{u,r}^{out}+S_{ul,4l} \phi_{u,l}^{out}.
\end{eqnarray}
Because of the negative norm of the $\phi_{ul}^{out}$ mode, conservation of the Bogoliubov norm imposes the modified unitarity condition $S^{\dagger}\eta\ S =S\eta\ S^{\dagger}$ with $\eta=diag( 1,1,-1)$ and the scattering matrix $S$ mixes positive and negative norm modes.

As a result, the Bogoliubov transformation \index{Bogoliubov transformation} relating the creation and destruction operators of the ``in" and ``out" scattering states is no longer trivial and mixes creation and destruction operators as follows
\begin{equation}
\label{Ssupsup}
     \left( \begin{array}{c}
       \hat a_{\omega}^{v,out} \\
       \hat a_{\omega}^{ur,out}  \\
       \hat a_{\omega}^{ul,out\dagger} \\
           \end{array} \right)
   = \left(
     \begin{array}{cccc}
       S_{vl,vr} & S_{vl,3l} & S_{vl,4l} \\
       S_{ur,vr} & S_{ur,3l} & S_{ur,4l} \\
       S_{ul,vr} & S_{ul,3l} & S_{ul,4l} \\
       \end{array}\right) \left(
                \begin{array}{c}
                  \hat a_{\omega}^{v,in}  \\
                  \hat a_{\omega}^{3in} \\
                  a_{\omega}^{4in\dagger}\\
                                  \end{array}
              \right).
\end{equation}
The non triviality of the Bogoliubov transformation has the crucial consequence that the ``in" and  ``out" vacua \index{vacuum state} no longer coincide $|0,in\rangle \neq   |0,out \rangle$: while the  $|0,in\rangle$ ``in" vacuum state (defined as the state annihilated by the  $\hat a_\omega^{(v,3,4),in}$ operators) contains no incoming phonons, it contains a finite amount of phonons in all three out-going modes due to a parametric conversion process taking place at the horizon,
\begin{eqnarray}\label{sund}
n_\omega^{u,r}&=&\langle 0,in|\hat a^{ur,out\dagger}_\omega\hat a^{ur,out}_\omega|0,in\rangle =|S_{ur,4l}|^2 ,\\ \label{sund1}
n_\omega^{u,l}&=&\langle 0,in|\hat a^{ul,out\dagger}_\omega\hat a^{ul,out}_\omega|0,in\rangle=|S_{ul,vr}|^2+|S_{ul,3l}|^2, \\ \label{sund2}
n_\omega^{v,l}&=&\langle 0,in|\hat a^{v,out\dagger}_\omega\hat a^{v,out}_\omega|0,in\rangle=|S_{vl,4l}|^2.
\end{eqnarray}
Note in particular the remarkable relation
\begin{equation}\label{sdod}
n_\omega^{u,l}=|S_{ul,vr}|^2+|S_{ul,3l}|^2=|S_{ur,4l}|^2+|S_{vl,4l}|^2=n_\omega^{u,r}+n_\omega^{v,l}.
\end{equation}
The physical meaning of the above relations can be understood as follows.
Suppose that at $t=-\infty$ we have prepared the system in the $|0,in\rangle$ vacuum state, so there are no incoming phonons. We are working in the Heisenberg  picture of Quantum Mechanics, so that $|0,in\rangle$ describes the state of our systems at all time. Now eqs. (\ref{sund}-\ref{sund2}) tell us that at late time in this state there will be outgoing quanta on both sides of the horizon: the vacuum has spontaneously emitted phonons. This occurs by converting vacuum fluctuation of the $k_4$  mode into real on shell Bogoliubov phonons (see lower left panel of Fig.\ref{fig:subsuper_inout})  in the hydrodynamic region.

While processes involving particle creation \index{particle creation} in time-varying settings are well-known in quantum mechanics, e.g. the dynamical Casimir effect \cite{dyncas,dyncas_exp}, the production of particles in a stationary background seems to contradict energy conservation \index{energy conservation}.
The solution of this puzzle relies in Eq.(\ref{sdod}): besides the positive energy $ur$ and $v$ phonons , there is also production of negative energy ($ul$) phonons, the so called \index{partners} ``partners" which propagate down in the supersonic region. The number of these latter equals the number of the formers. This is how energy conservation \index{energy conservation} and particles production coexists in our stationary systems. As particles are produced in pairs with opposite $\pm\omega$ frequencies, energy is conserved.

Now let us give a closer look at  Eq. (\ref{sund}): according to this, an hypothetical observer sitting far away from the horizon in the  subsonic  region at $x\to +\infty$ will reveal a flux of phonons coming from the horizon. This is just the analogue of Hawking black hole radiation \index{Hawking radiation}. The number of phonons of this kind emitted per unit time and per unit bandwidth is
\begin{equation}\label{ppdr}
\frac{dN_\omega^{u,r}}{dt\,d\omega}=|S_{ur,4l}|^2\simeq \frac{(c_r+v_0)}{(c_r-v_0)}\frac{(v_0^2-c_l^2)^{3/2}}{(c_r^2-c_l^2)}
\frac{2c_r}{c_l\xi_l \omega}  \ .
\end{equation}
The $\frac{1}{\omega}$ behavior of the above expression is reminiscent of the low frequency expansion of a thermal Bose distribution \cite{rpc}
\begin{equation}
n_T(\omega)=\frac{1}{e^{\frac{k_BT}{\hbar \omega}}-1}\simeq \frac{k_B T}{\hbar \omega} +...
\end{equation}
and one can try to identify the $\frac{1}{\omega}$ coefficient of (\ref{ppdr}) as an effective temperature
\begin{equation}
T=\frac{\hbar}{k_B} \frac{(c_r+v_0)}{(c_r-v_0)}\frac{(v_0^2-c_l^2)^{3/2}}{(c_r^2-c_l^2)}
\frac{2c_r}{c_l\xi_l }\ .
\end{equation}
As the surface gravity \index{surface gravity} of our toy model with an abrupt discontinuity in the flow is formally infinite while the temperature remains finite, the connection of the analog model to the original gravitational framework seems to fail. However, to investigate the correspondence with the gravitational black holes, one has to consider more general and realistic velocity profiles where the transition from the subsonic region to the supersonic one is smooth enough to justify the hydrodynamical approximation. Accurate numerical calculations in this regime show that the emission is indeed thermal in this case and the temperature is to a good accuracy determined by the surface gravity $\kappa$ of the associated black hole according to Eq. (\ref{hrsh}). As it was shown in~\cite{machpare, finapare}, the original Hawking's prediction for the emission temperature holds provided the spatial variation of the flow parameters occurs on a characteristic length scale longer than $\xi^{2/3}\kappa^{-1/3}$. Of course, the thermal spectrum is restricted to frequencies lower than the upper cut-off at $\omega_{max}$: above this frequency, one in fact recovers the physics of the sub-sub interface where no emission takes place. These results, together with the full numerical simulation of~\cite{numerics} confirm that the emission of Hawking radiation is not an artifact of the hydrodynamical approximation and provide an independent validation of the model of phonon propagation based on the metric Eq.(\ref{acm}).

Even in the most favorable configurations, realistic estimates of the Hawking  temperature in atomic BECs give values of the order of $10\,$nK, that is one order of magnitude lower than the typical temperature of the condensates ($100\,$nK). This makes the Hawking emission of Bogoliubov phonons in BECs a quite difficult effect to reveal in an actual experiment, as the interesting signal is masked by an overwhelming thermal noise.

A proposal to overcome this difficulty was put forward in~\cite{lettera}: as the pairs of Bogoliubov excitations produced by the Hawking process originate from the same vacuum fluctuation, their strong correlation is expected to be responsible for specific features in the correlation function of density fluctuations. This idea was soon confirmed by numerical simulations of the dynamics of atomic condensates in acoustic black hole  configurations.
The features analytically predicted in~\cite{lettera} are indeed visible in the density correlation pattern and, moreover, are robust with respect to a finite temperature.
At present, this method represents the most promising strategy to experimentally detect the analog Hawking effect in atomic Bose-Einstein condensates \index{Bose-Einstein condensates}. A first investigation of the power of density correlation techniques in the context of an analog dynamical Casimir effect in condensates has been recently reported in~\cite{dyncas_BEC_exp} along the lines of the theoretical proposal in~\cite{dyncas_BEC}.

\subsection{Density-density correlations \index{density-density correlations}}
\begin{figure}[htbp]
\begin{center}
\includegraphics[width=12cm,clip]{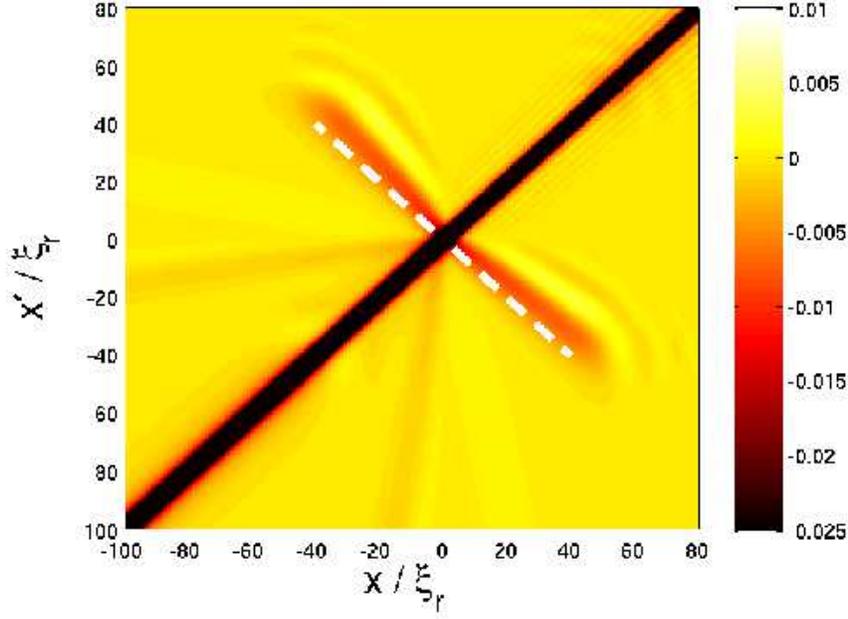}
\end{center}
%
%
\caption{Color plots of the rescaled density correlation function $(n_0\xi_r)\times[G^{(2)}(x,x')-1]$ a time $g_r n t/\hbar=160$ after the switch-on of the black hole horizon. The calculation has been performed using the truncated-Wigner method of \cite{numerics}. Black hole parameters: $|v_0|/c_l=1.5$, $|v_0|/c_r=0.75$. The dashed white line indicates the analytically expected position (\ref{hawpea}) of the negative peak in the density correlation signal. 
}
\label{fig:numeric}       
\end{figure}

The fact that the density correlation function in BECs exhibits characteristic peaks associated to the phonons creation \`a la Hawking  can be easily seen in our simple toy model in the following way. For simplicity, let us restrict our attention to the contribution to the density-density correlation function due to the ``out" particles
and  consider the decomposition
\begin{multline}
\hat n^1(t,x)\simeq n\int_{0}^{\omega_{max}}\left[\hat a_{\omega}^{v,out}(\phi_{v,l}^{out}+\varphi_{v,l}^{out})\right.+
+\hat a_{\omega}^{ur,out}(\phi_{u,r}^{out}+\varphi_{u,r}^{out})+\\
+\left.\hat a_{\omega}^{ul,out\dagger}(\phi_{u,l}^{out}+\varphi_{u,l}^{out})+h.c.\right].
\end{multline}
Expanding the ``out" creation and annihilation operators in terms of the ``in" ones and using the approximate form of the $S$ matrix elements given by Eq. (\ref{Ssupsup}), and finally evaluating expectation values on the $|0,in\rangle$ ``in" vacuum state, one finds that the above expression describes correlations between the ($ur$) and ($ul$) particles and the ($ur$) and ($vl$) particles if the points $x$ and $x'$ are taken on opposite sides with respect the horizon, while one finds ($ul$)-($vl$) correlations if both points are inside the horizon. If both $x,x'$ are located outside the horizon, correlations just show a monotonic decrease with distance as in the sub-sub case.

As in general one has $|S_{vl,4l}|\ll |S_{ul,4l}|$, the main contribution to the density correlation in the $x<0$ and $x'<0$ sector comes from the ($ul$)-($ur$) term describing correlations between the Hawking phonon ($ur$) and its partner \index{partner} ($ul$). Integrating over all frequencies upto $\omega_{max}$, one obtains term of the form~\cite{rpc}
\begin{equation}
\label{correlation_spatial}
G^{(2)}(x,x')\sim-\frac{1}{4\pi n}\frac{(v_0^2-c_l^2)^{3/2}}{c_l(v_0+c_l)(v_0-c_r)(c_r-c_l)}\frac{\sin\left[\omega_{max}(\frac{x'}{v_0+c_r}-\frac{x}{v_0+c_l})\right]}{\frac{x'}{v_0+c_r}-\frac{x}{v_0+c_l}}\ .
\end{equation}
From this expression, it is easy to see that the density-density correlation function has a  negative value and is peaked along the half-line
\begin{equation}\label{hawpea}
\frac{x'}{v_0+c_r}=\frac{x}{v_0+c_l}.
\end{equation}
The stationarity \index{stationarity} of the Hawking process is apparent in the fact that the peak value of Eq.(\ref{correlation_spatial}) does not depend on the distance from the horizon.

The physical picture that emerges from this mathematical derivation is that pairs of ($ul$) and ($ur$) phonons are continuously created by the horizon at each time $t$ and then propagate on opposite directions at speeds $v_{ul}= v_0+c_l<0$ for ($ul$) and $v_{ur}=v_0 +c_r>0$ for ($ur$). At time $\Delta t$ after their emission they are located at $x= v_{ul}\,\Delta t$ and $x'=v_{ur}\,\Delta t$, which explains the geometrical shape of the peak line Eq. (\ref{hawpea}) where correlations are strongest. An example of numerical picture of the correlation function of density fluctuations is shown in Fig.\ref{fig:numeric}: the dashed line indicate the expected position of the peak line Eq.(\ref{hawpea}). A detailed discussion of the other peaks (that are barely visible on the color scale of the figure) can be found in~\cite{numerics,rpc}.

\subsection{Remarks}

Let us try to summarize the results discussed in the present chapter.
We have seen that if a stationary flowing BEC shows a horizon-like boundary separating an upstream subsonic region from a downstream supersonic one, a spontaneous emission of Bogoliubov phonons occurs at the horizon by converting zero-point quantum fluctuations \index{zero-point quantum fluctuations} into real and observable radiation quanta. The emitted radiation appears to an observer outside the horizon in the subsonic region to have an approximately thermal distribution: this is the analog Hawking effect in BECs.

As in a gravitational context nothing can travel faster than light, the horizon has a well-defined meaning of surface of no return: no physical signal can travel from inside the black hole to the outside crossing the horizon in the outward direction.
In the case of an atomic BECs, the ``sonic" horizon is defined as the surface where the speed of sound $c$ equals the velocity $|v_0|$ of the fluid: for what concerns the hydrodynamical $u,\ v$ modes at low wavevector, the acoustic black hole  exactly mimics what happens in gravity: no long wavevector sound wave can cross the horizon in the upstream direction.
  On the other hand the dispersion of Bogoliubov modes  in an atomic BEC shows significant super-luminal corrections: the higher the wavevector of the excitation, the larger its group velocity. As a result, the $k_{3,4}$ modes are able to travel in the upstream direction in the super-sonic region inside the horizon and therefore to escape from the black hole. In contrast to the hydrodynamic modes, they are not trapped inside and do not see any horizon: for them the gravitational analogy \index{gravitational analogy} has no meaning.

Some authors have recently introduced wavevector dependent {\em rainbow metrics} to describe the propagation of different modes at different wave vectors and have defined several distinct concepts of horizon, such as the {\em phase horizon} and the {\em group horizon}. Our opinion is that these additional concepts may end up hiding the essence of Hawking radiation behind unessential details.

The key ingredient in order to have the emission of radiation in the ``in" vacuum state is in fact the presence of negative energy states \index{negative energy states} that allow to emit a pair of quanta while conserving energy: this requires that the flow undergoes supersonic motion in some spatial region. The main role of a horizon where the flow goes from sub- to super-sonic is to determine the thermal shape spectral distribution of the emitted radiation.
To better appreciate this fundamental point, the next chapter will be devoted to a short discussion of configurations with a super-sonic flow on both sides of the interface: as it was first pointed out in~\cite{finazzi_supersuper}, a spontaneous emission of radiation takes place in this case in spite of the total absence of a horizon: sound waves are always dragged by the super-sonic flow and can not propagate upstream. Because of the absence of an horizon, the resulting spectral distribution of the associated zero-point radiation is however very different from the thermal Hawking radiation, with a low-frequency tail dominated by a constant term  instead of $1/\omega$.

\section{Supersonic-supersonic configuration}
\label{sec:7}

\begin{figure}[htbp]
\begin{center}
\includegraphics[width=10cm,clip]{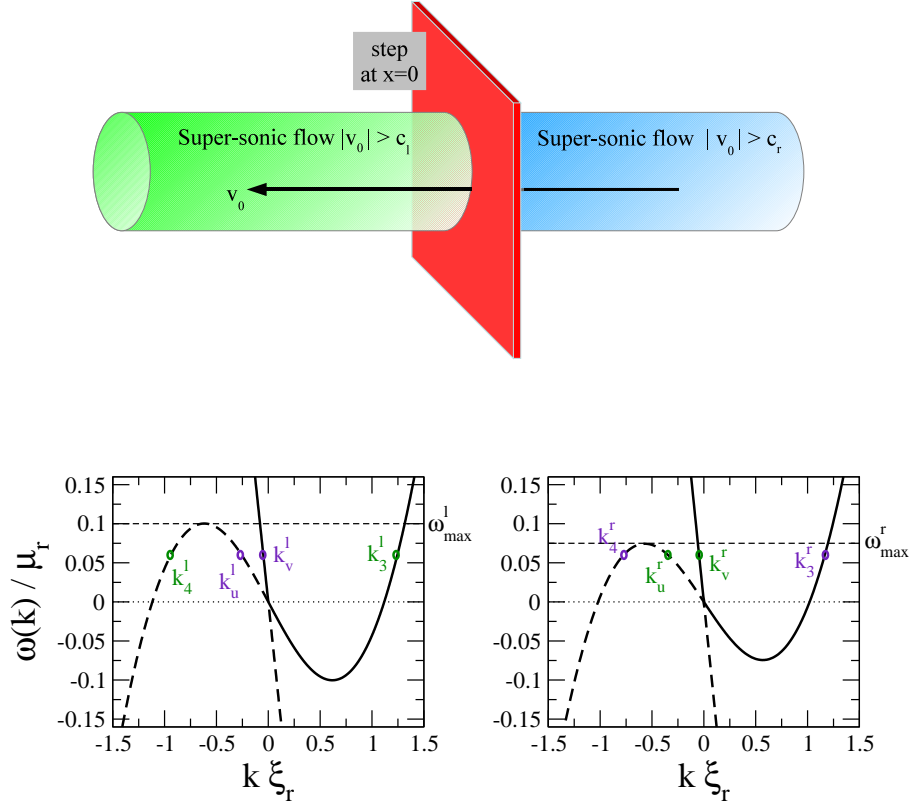}
\includegraphics[width=12cm,clip]{figura_dispersion_supersuper.eps}
\end{center}
%
%
\caption{Upper panel: sketch of the supersonic-supersonic flow configuration. Low panels: dispersion relation of Bogoliubov modes  in the asymptotic regions away from the horizon. }
\label{fig:supersuper_flow}       
\end{figure}

Consider a BEC undergoing an everywhere supersonic motion \index{supersonic flow} , with a sound velocity profile varying abruptly at $x=0$: no sonic horizon is present in this setting and at all points long wavelength sound waves are dragged in the downstream direction by the underlying flowing fluid.

The dispersion relation  pattern on either sides of the discontinuity is shown in the lower panels of Fig.\ref{fig:supersuper_flow}. For $\omega<\omega_{max}=\textrm{min}[\omega_{max}^{l}, \omega_{max}^{r}]$, one has  four oscillatory solutions in both regions with real wavevectors.  The $k_u$ and $k_v$ hydrodynamic solutions propagate in the downstream direction (i.e. to the left with negative $v_0$) while the large wavevector $k_3$ and $k_4$ solutions are able to propagate upstream. While the $k_{v,3}$ solutions correspond to positive norm modes, the $k_{u,4}$ are negative norm ones.

The general solution of the mode equations in both regions reads
\begin{multline}
\phi_\omega^{r(l)}=e^{-i\omega t}\left[A_v^{l(r)}D_v^{l(r)}e^{ik_v^{l(r)}x}+A_u^{l(r)}D_u^{l(r)}e^{ik_u^{l(r)}x}+ \right.\\
+\left.A_3^{l(r)}D_3^{l(r)}e^{ik_3^{l(r)}x}+A_4^{l(r)}D_4^{l(r)}e^{ik_4^{l(r)}x}\right]\ .
\end{multline}
As usual, the left and right amplitudes are related by
\begin{equation}
     \left( \begin{array}{c}
       A_v^l \\
       A_u^l \\
       A_3^l \\
       A_4^l \\
     \end{array} \right)
   =M \left(
                \begin{array}{c}
                             A_v^r \\
       A_u^r \\
       A_3^r \\
       A_4^r \\
                \end{array}
              \right),
\end{equation}
the matching matrix being given by $M=W_l^{-1}W_r$ with
\begin{equation}
\label{eq:wlass}
W_{l(r)}=\left(
     \begin{array}{cccc}
       D_v^{l(r)} & D_u^{l(r)} & D_{3}^{l(r)} & D_{4}^{l(r)}\\
       ik_v^{l(r)}D_v^{l(r)} & ik_u^{l(r)}D_u^{l(r)} & ik_3^{l(r)}D_{3}^{l(r)} & ik_4^{l(r)}D_{4}^{l(r)} \\
       E_v^{l(r)} & E_u^{l(r)} & E_{3}^{l(r)} & E_{4}^{l(r)} \\
       ik_v^{l(r)}E_v^{l(r)} & ik_u^{l(r)}E_u^{l(r)} & ik_3^{l(r)}D_{3}^{l(r)} & ik_4^{l(r)}D_{4}^{l(r)}  \\
\end{array}\right).
\end{equation}

\begin{figure}[htbp]
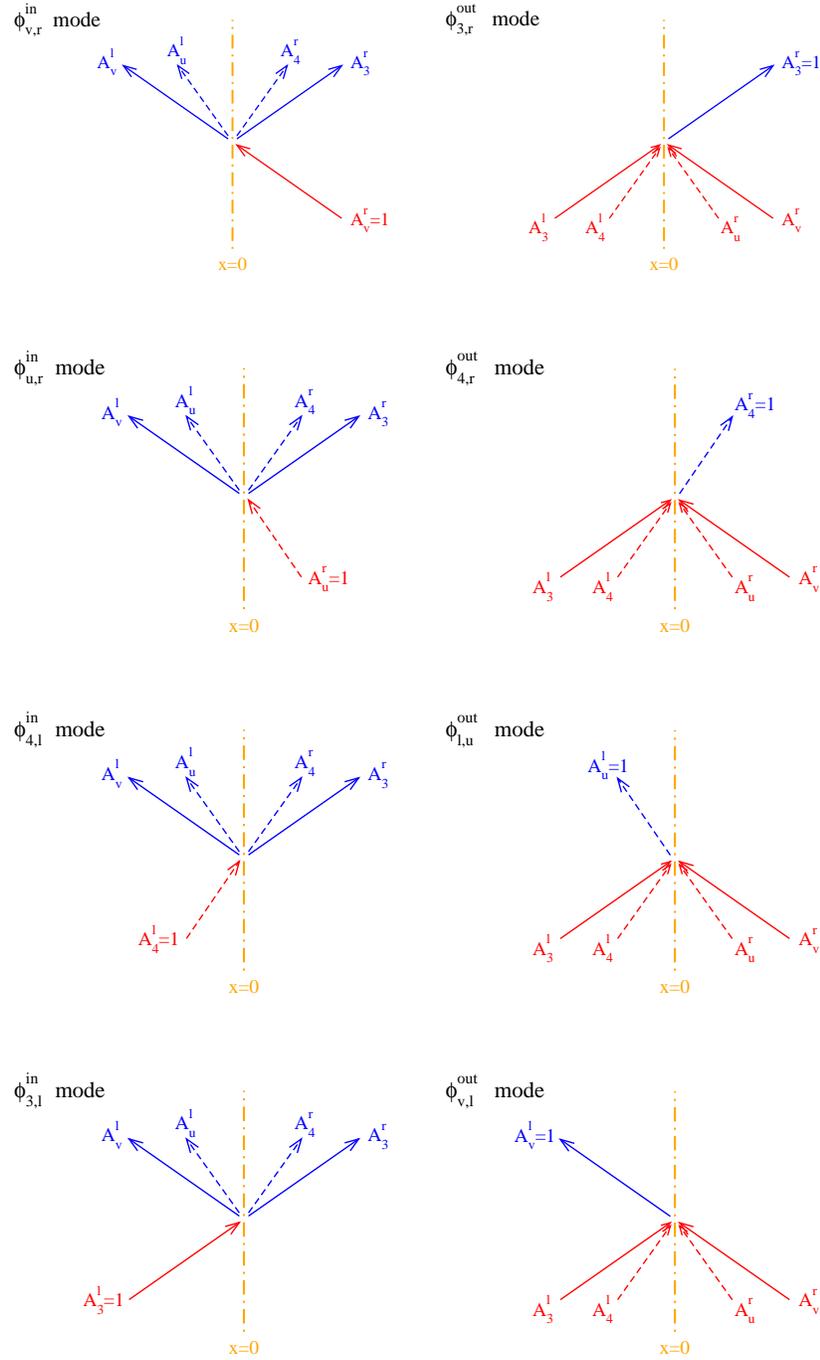

\begin{center}
\includegraphics[width=5.cm,clip]{figura_modes3a.eps}
\hspace{0.5cm}
\includegraphics[width=5.0cm,clip]{figura_modes3e.eps} \\
\vspace*{1cm}
\includegraphics[width=5.0cm,clip]{figura_modes3b.eps}
\hspace{0.5cm}
\includegraphics[width=5.0cm,clip]{figura_modes3f.eps}\\
\vspace*{1cm}
\includegraphics[width=5.0cm,clip]{figura_modes3c.eps}
\hspace{0.5cm}
\includegraphics[width=5.0cm,clip]{figura_modes3g.eps}\\
\vspace*{1cm}
\includegraphics[width=5.0cm,clip]{figura_modes3d.eps}
\hspace{0.5cm}
\includegraphics[width=5.0cm,clip]{figura_modes3h.eps}
\end{center}
%
%
\caption{Sketch of the Bogoliubov modes involved in the ``in'' (left panels) and ``out'' (right panels) basis. The mode labels refer to the dispersion shown in the lower panels of Fig.\ref{fig:supersuper_flow}. }
\label{fig:supersuperinout}       
\end{figure}

As sketched in Fig.\ref{fig:supersuperinout}, the ``in" basis is here defined by four in-going waves: two of them  ($u,v$) are incident on the discontinuity from the right (left panels on the first and second rows); the two others  ($3,4$) are incoming from the left (left panels on the third and fourth rows). The ``out" basis is defined along the same lines as sketched in the four panels of the right column.

The field operator can be expanded either in the ``in" or in the ``out" basis as
\begin{multline}
\hat\phi=\int_{0}^{\omega_{max}}d\omega\left[\hat a_{\omega}^{v,in}\phi_{v,r}^{v}+
\hat a_{\omega}^{u,in\dagger}\phi_{u,r}^{in}\right.+
\hat a_{\omega}^{3,in}\phi_{3,l}^{in}+\hat a_{\omega}^{4,in\dagger}\phi_{4,l}^{in}+ \\ +\left.\hat a_{\omega}^{v,in\dagger}\varphi_{v,r}^{in*}+ a_{\omega}^{u,in}\varphi_{u,r}^{in*}+ \hat a_{\omega}^{3,in\dagger}\varphi_{3,l}^{in*}+\hat a_{\omega}^{4,in}\varphi_{4,l}^{in*}\right]
\end{multline}
or
\begin{multline}
\hat\phi=\int_{0}^{\omega_{max}}d\omega\left[\hat a_{\omega}^{v,out}\phi_{v,l}^{out}+
\hat a_{\omega}^{u,out\dagger}\phi_{u,l}^{out}+
\hat a_{\omega}^{3,out}\phi_{3,r}^{out}+\hat a_{\omega}^{4,out\dagger}\phi_{4,r}^{out}\right.+\\
+\left. \hat a_{\omega}^{v,out\dagger}\varphi_{v,l}^{out*}+\hat a_{\omega}^{u,out}\varphi_{u,l}^{out*} + \hat a_{\omega}^{3,out\dagger}\varphi_{3,r}^{out*}+\hat a_{\omega}^{4,out}\varphi_{4,r}^{out*}\right].
\end{multline}
The relation between the ``in" and ``out" basis are
\begin{eqnarray}\label{eq:outinaass}
  \phi_{v,r}^{in}&=& S_{vl,vr} \phi_{v,l}^{out}+S_{ul,vr} \phi_{u,l}^{out}+S_{3r,vr} \phi_{3,r}^{out}
+S_{4r,vr} \phi_{4,r}^{out} \ , \\
 \phi_{u,r}^{in}&=& S_{vl,ur} \phi_{v,l}^{out}+S_{ul,ur} \phi_{u,l}^{out}+S_{3r,ur} \phi_{3,r}^{out}
+S_{4r,ur} \phi_{4,r}^{out} \ , \\
  \phi_{3,l}^{in} &=& S_{vl,3l} \phi_{v,l}^{out}+S_{ul,3l} \phi_{u,l}^{out}+S_{3r,3l} \phi_{3,r}^{out}
+S_{4r,3l} \phi_{4,r}^{out} \ , \\
  \phi_{4,l}^{in} &=& S_{vl,4l} \phi_{v,l}^{out}+S_{ul,4l} \phi_{u,l}^{out}+S_{3r,4l} \phi_{3,r}^{out}
+S_{4r,4l} \phi_{4,r}^{out} \
\end{eqnarray}
and the corresponding relation between the ``in" and ``out" annihilation and creation operators \index{Bogoliubov transformation} \index{scattering matrix} reads
\begin{equation}
\label{Ssupsups}
     \left( \begin{array}{c}
       \hat a_{\omega}^{v,out} \\
       \hat a_{\omega}^{u,out\dagger}  \\
       \hat a_{\omega}^{3,out} \\
        \hat a_{\omega}^{4,out\dagger} \\
           \end{array} \right)
   = \left(
     \begin{array}{cccc}
       S_{vl,vr} & S_{vl,ur} & S_{vl,3l} & S_{vl,4l}\\
       S_{ul,vr} & S_{ul,ur} & S_{ul,3l} & S_{ul,4l}\\
       S_{3r,vr} & S_{3r,ur} & S_{3r,3l} & S_{3r,4l}\\
       S_{4r,vr} & S_{4r,ur} & S_{4r,3l} & S_{4r,4l}\\
       \end{array}\right) \left(
                \begin{array}{c}
                  \hat a_{\omega}^{v,in}  \\
                  \hat a_{\omega}^{u,in\dagger}  \\
                  \hat a_{\omega}^{3,in} \\
                  a_{\omega}^{4,in\dagger}\\
                                  \end{array}
              \right).
\end{equation}
Explicit expressions for the corresponding $16$ amplitudes are  listed in the Appendix.

Here we see again that the $S$ matrix mixes creation and annihilation operators. As a consequence, the $|0,in\rangle$   and  $|0,out\rangle$ vacua do not coincide: in particular the ``in" vacuum $|0,in\rangle$ state with no incident quanta leads to a finite amount of out-going particles that can be detected: in the left region, they belong to the $u,\ v$ modes; in the right region, they belong to the $3,\ 4$ modes.

More precisely
\begin{eqnarray}\label{sundss}
n_\omega^{v,l}&=&\langle 0,in|\hat a^{v,out\dagger}_\omega\hat a^{v,out}_\omega|0,in\rangle =|S_{vl,ur}|^2+
|S_{vl,4l}|^2 \\
n_\omega^{u,l}&=&\langle 0,in|\hat a^{u,out\dagger}_\omega\hat a^{u,out}_\omega|0,in\rangle=|S_{ul,vr}|^2+|S_{ul,3l}|^2 \\
n_\omega^{3,r}&=&\langle 0,in|\hat a^{3,out\dagger}_\omega\hat a^{3,out}_\omega|0,in\rangle=|S_{3r,ur}|^2+|S_{3r,4l}|^2 \\
n_\omega^{4,r}&=&\langle 0,in|\hat a^{4,out\dagger}_\omega\hat a^{4,out}_\omega|0,in\rangle=|S_{4r,vr}|^2+|S_{4r,3l}|^2
\end{eqnarray}
and unitarity of the $S$ matrix imposes that
\begin{equation}
n_\omega^{v,l}+n_\omega^{3,r}=n_\omega^{u,l}+n_\omega^{4,r}:
\end{equation}
the number of positive energy particles equals the number of negative energy ones.
From the explicit expressions for the $S$ matrix elements listed in the Appendix, it is immediate to see that all spectral distributions $n_\omega^{v,l}$, $n_\omega^{u,l}$, $n_\omega^{3,r}$ and $n_\omega^{4,r}$ at low frequencies are dominated by constant terms, in stark contrast with the $1/\omega$ shape of the thermal Hawking radiation.

\section{Conclusions}
\label{sec:8}


In this chapter, we have given an introductory review to Hawking radiation effects in atomic Bose-Einstein condensates Bose-Einstein condensates. Focussing our attention on a simple toy model based on a piecewise uniform flow interrupted by sharp interfaces, we have made use of the standard Bogoliubov theory Bogoliubov theory of dilute condensates to obtain analytical predictions for the quantum vacuum emission of phonons that is emitted by the interface: necessary and sufficient condition for this emission to occur is that the flow be somewhere super-sonic. While the low-frequency part of the emission follows an approximately thermal form for a black-hole interface separating a sub-sonic upstream region from a super-sonic downstream one, a completely different spectrum is found for flows that do not show any horizon and are everywhere super-sonic.

The interest of our development is manyfold: on one hand, our analytical treatment provides an intuitive understanding of Hawking radiation based on a Bogoliubov generalization of the scattering of waves by square potentials in one dimensional Schr\"odinger equation. On the other hand, our derivation is however completely ``ab initio", based on the fundamental microscopic quantum description of the BEC without any recourse to the gravitational analogy. As a result, it does not depend on the hydrodynamic approximation that underlies the introduction of the effective metric and shows that the Hawking effect in atomic BECs is not at all an artifact of the low wavelength (hydrodynamical) approximation: no transplanckian problem \index{transplanckian problem} is present which may cast doubts on the derivation, rather our derivation shows that the transplanckian problem is itself an artifact of the hydrodynamical approximation.

The intense theoretical and experimental activity that is currently in progress makes us confident that the existence of analog Hawking radiation will be soon experimentally confirmed. The robustness of Hawking radiation with respect to the microscopic details of the condensed-matter system would be a strong indication that, in spite we do not have any knowledge of the quantum microscopic description of gravity, Hawking's prediction of black hole radiation with its important thermodynamical implications is a real milestone in our understanding of Nature.

\begin{acknowledgement}
Continuous stimulating discussions with S. Finazzi, R. Parentani and N. Pavloff are warmly acknowledged. IC acknowledges partial financial support from ERC via the QGBE grant.
\end{acknowledgement}

\section*{Appendix}
\addcontentsline{toc}{section}{Appendix}

In this Appendix, we give the explicit expressions for the leading order in the small $\omega$ limit of the $S$ Matrix coefficients for the supersonic-supersonic configuration  treated in Sect.~\ref{sec:7}. Note in particular how the ones involved in the vacuum emission ($S_{vl,ur}$, $S_{ul,vr}$,
$S_{3r,4l}$, $S_{4r,3l}$) 
grow as $\sqrt{\omega}$ at low $\omega$, while
($S_{4r,vr}$, $S_{3r,ur}$, $S_{ul,3l}$, $S_{vl,4l}$) tend to constant.
\begin{eqnarray}\label{smatsupsup}
S_{vl,vr}&=&\sqrt{\frac{v_0+c_l}{v_0-c_l}}
\frac{(c_r^2-c_l^2)\sqrt{\omega\xi_l}}{2\sqrt{2}(v_0^2-c_l^2)^{1/4}(v_0^2-c_r^2)}\ ,\nonumber \\
S_{ul,vr} &=& \sqrt{\frac{v_0-c_l}{v_0+c_l}}
\frac{(c_r^2-c_l^2)\sqrt{\omega\xi_l}}{2\sqrt{2}(v_0^2-c_l^2)^{1/4}(v_0^2-c_r^2)}\ ,\nonumber \\
S_{3r,vr} &=& \frac{\sqrt{v_0^2-c_l^2}+\sqrt{v_0^2-c_r^2}}{2(v_0^2-c_l^2)^{1/4}(v_0^2-c_r^2)^{1/4}}\ ,\nonumber \\
S_{4r,vr} &=& \frac{\sqrt{v_0^2-c_l^2}-\sqrt{v_0^2-c_r^2}}{2(v_0^2-c_l^2)^{1/4}(v_0^2-c_r^2)^{1/4}}\ ,\nonumber \\
S_{vl,ur}&=&\sqrt{\frac{v_0+c_l}{v_0-c_l}}
\frac{(c_l^2-c_r^2)\sqrt{\omega\xi_l}}{2\sqrt{2}(v_0^2-c_l^2)^{1/4}(v_0^2-c_r^2)}\ ,\nonumber \\
S_{ul,ur} &=& \sqrt{\frac{v_0-c_l}{v_0+c_l}}
\frac{(c_r^2-c_l^2)\sqrt{\omega\xi_l}}{2\sqrt{2}(v_0^2-c_l^2)^{1/4}(v_0^2-c_r^2)}\ ,\nonumber \\
S_{3r,ur} &=& \frac{\sqrt{v_0^2-c_r^2}-\sqrt{v_0^2-c_l^2}}{2(v_0^2-c_l^2)^{1/4}(v_0^2-c_r^2)^{1/4}}\ ,\nonumber \\
S_{4r,ur} &=& \frac{\sqrt{v_0^2-c_l^2}+\sqrt{v_0^2-c_r^2}}{2(v_0^2-c_l^2)^{1/4}(v_0^2-c_r^2)^{1/4}}\ ,\nonumber \\
S_{vl,3l} &=& \frac{c_l+c_r}{2\sqrt{c_lc_r}},\nonumber \\
S_{ul,3l} &=& i\frac{c_l-c_r}{2\sqrt{c_lc_r}},\nonumber \\
S_{3r,3l} &=& \frac{(v_0^2-c_r^2)^{1/4}(c_l^2-c_r^2)\sqrt{c_l\xi_l\omega}}{2\sqrt{2c_r}(c_r-v_0)(v_0^2-c_l^2)}
,\nonumber \\
S_{4r,3l} &=& \frac{(v_0^2-c_r^2)^{1/4}(c_l^2-c_r^2)\sqrt{c_l\xi_l\omega}}{2\sqrt{2c_r}(c_r-v_0)(v_0^2-c_l^2)}
,\nonumber \\
S_{vl,4l} &=& \frac{c_l-c_r}{2\sqrt{c_lc_r}},\nonumber \\
S_{ul,4l} &=& i\frac{c_l+c_r}{2\sqrt{c_lc_r}},\nonumber \\
S_{3r,4l} &=& \frac{(v_0^2-c_r^2)^{1/4}(c_l^2-c_r^2)\sqrt{c_l\xi_l\omega}}{2\sqrt{2c_r}(c_r+v_0)(c_l^2-v_0^2)}
,\nonumber \\
S_{4r,4l} &=& i\frac{(v_0^2-c_r^2)^{1/4}(c_r^2-c_l^2)\sqrt{c_l\xi_l\omega}}{2\sqrt{2c_r}(c_r+v_0)(c_l^2-v_0^2)}\ .
\end{eqnarray}

\end{document}